\documentclass[prd,twocolumn,showpacs,groupedaddress]{revtex4}
\newif\ifdraft 
\draftfalse		
\ifdraft
  \newcommand{\PC}[1]{{\bf [Predrag: #1]}}
  \newcommand{\ADK}[1]{{\bf [Tony: #1]}}
  \newcommand{\HE}[1]{{\bf [Henry: #1]}}
\else
  \newcommand{\PC}[1]{}
  \newcommand{\HE}[1]{}
  \newcommand{\ADK}[1]{}
\fi
\newif\ifpdf			
\ifx\pdfoutput\undefined\pdffalse\else\pdfoutput=1\pdftrue\fi
\ifpdf\pdfcompresslevel=9\fi	
\ifpdf\usepackage[backref]{hyperref}\fi
\ifpdf\let\myhref=\href\def\href#1#2{\penalty-20\myhref{#1}{\tt #2}}%
  \else\def\href#1#2{{\penalty-20\tt #2}}\fi
\usepackage{graphicx}
\ifpdf\fi
\ifpdf\else\usepackage[dvips]{color}\fi     
\newdimen\onebox
\newdimen\boxsize
\newcount\boxnum
\gdef\mult#1#2#3{
  \ifx#1\relax\else\ifx#2\relax\else %
    #1=#2\ifx#3\relax\else\multiply#1#3\fi\fi\fi}
\ifpdf\newcommand{\btrackYt}[2]{%
  \boxnum=#2%
  \onebox=8pt%
  \mult{\boxsize}{\onebox}{\boxnum}%
  \parbox{\boxsize}{\includegraphics[width=\boxsize]{figs/f_#1.pdf}}}%
\else\newcommand{\btrackYt}[2]{%
  \boxnum=#2%
  \onebox=8pt%
  \mult{\boxsize}{\onebox}{\boxnum}%
  \parbox{\boxsize}{\includegraphics[width=\boxsize]{figs/f_#1.eps}}}%
\fi
\ifpdf%
\else%
\fi
\ifpdf\newcommand{\btrackYb}[2]{%
  \boxnum=#2%
  \onebox=5pt%
  \mult{\boxsize}{\onebox}{\boxnum}%
  \parbox{\boxsize}{\includegraphics[width=\boxsize]{figs/f_#1.pdf}}}%
\else\newcommand{\btrackYb}[2]{%
  \boxnum=#2%
  \onebox=5pt%
  \mult{\boxsize}{\onebox}{\boxnum}%
  \parbox{\boxsize}{\includegraphics[width=\boxsize]{figs/f_#1.eps}}}%
\fi
\newcommand{\rf}      [1] {~\cite{#1}}
\newcommand{\refref}  [1] {ref.~\cite{#1}}

\newcommand{\refrefs} [1] {refs.~\cite{#1}}

\newcommand{\refeq}   [1] {(\ref{#1})}

\newcommand{\reffig}  [1] {fig.~\ref{#1}}

\newcommand{\refsect} [1] {\S\ref{#1}}

\newcommand{\refappe} [1] {appendix~\ref{#1}}

\newcommand{\YoungPOp}{Young projec\-tion opera\-tor} 
\newcommand{\YoungPOps}{Young projec\-tion opera\-tors}
\newcommand{\irrep}{irrep} 
\newcommand{\tr}{\mathop{\rm tr}}
\newcommand{\numBoxes}{k} 
\newcommand{\sign}{\mathop{\rm sgn}}
\newcommand{\sun}{{\(SU(n)\)}}
\newcommand{\un}{{\(U(n)\)}}
\newcommand{\rom}[1]{{\mathrm{#1}}}
\newcommand{\X}{\rom{X}} 
\newcommand{\Y}{\rom{Y}}
\newcommand{\Z}{\rom{Z}}
\newcommand{\U}{\rom{U}}
\newcommand{\V}{\rom{V}}
\newcommand{\W}{\rom{W}}
\newcommand{\symgrp}{{\mathcal S}} 
\newcommand{\row}{\rom{S}} 
\newcommand{\col}{\rom{A}} 
\newcommand{\theconst}{M}
\newcommand{\threenj}{{\(3n\)-\(j\)}} 
\newcommand{\C}{{\mathbb C}} 
\newcommand{\id}{{\mathbb I}} 
   
\usepackage{amssymb}
\begin{document}

\title{Diagrammatic Young Projection Operators for \un}

\author{Henriette Elvang$\,^{1}$ } 
\author{Predrag Cvitanovi\'c$\,^{2}$} 
\author{Anthony D. Kennedy$\,^{3}$} 
\affiliation{
  $^1$ Department of Physics, UCSB, Santa Barbara, CA 93106 \\
  $^2$ School of Physics, Georgia Institute of Technology, 
    Atlanta, GA 30332-0430 \\
  $^3$ School of Physics, JCMB, King's Buildings, University of Edinburgh,
    Edinburgh EH9 3JZ, Scotland
}
\date{\today}
\begin{abstract}
  \noindent We utilize a diagrammatic notation for invariant tensors to
  construct the \YoungPOps\ for the irreducible representations of the unitary
  group \un, prove their uniqueness, idempotency, and orthogonality, and
  rederive the formula for their dimensions. We show that all \un\ invariant
  scalars (\threenj\ coefficients) can be constructed and evaluated
  diagrammatically from these \un\ \YoungPOps. We prove that the values of all
  \un\ \threenj\ coefficients are proportional to the dimension of the maximal
  representation in the coefficient, with the proportionality factor fully
  determined by its \(\symgrp_\numBoxes\) symmetric group value. We also derive
  a family of new sum rules for the 3-\(j\) and 6-\(j\) coefficients, and
  discuss relations that follow from the negative dimensionality
  theorem.\parfillskip=0pt\par
\end{abstract}
\maketitle

\section{Introduction} \label{s-intro}

Symmetries are beautiful, and theoretical physics is replete with them, but
there comes a time when a calculation must be done. Innumerable calculations in
high-energy physics, nuclear physics, atomic physics, and quantum chemistry
require construction of irreducible many-particle states (\irrep s),
decomposition of Kronecker products of such states into \irrep s, and
evaluations of group theoretical weights (Wigner \threenj\ symbols, reduced
matrix elements, quantum field theory ``vacuum bubbles''). At such times
effective calculational methods gain in appreciation.

In his 1841 fundamental paper\rf{Jacobi1841} on the determinants today known as
``Jacobians'', Jacobi initiated the theory of \irrep s of the symmetric group
\(\symgrp_\numBoxes\). Schur used the \(\symgrp_\numBoxes\) \irrep s to develop
the representation theory of \(GL(n;\C)\) in his 1901
dissertation\rf{Schur1901}, and already by 1903 the Young
tableaux\rf{Young1977,WolframYoung} came into use as a powerful tool for
reduction of both \(\symgrp_\numBoxes\) and \(GL(n;\C)\) representations. In
quantum theory the group of choice\rf{Weyl31} is the unitary group \un, rather
than the general linear group \(GL(n;\C)\). Today this theory forms the core of
the representation theory of both discrete and continuous groups, described in
many excellent textbooks\rf{HAMERMESH,Messiah62,Coleman68,LICHTENBERG,%
Wybourne70,Georgi99,Boerner70,Andrews76,Stanley99,Fulton91,Fulton99,Manivel01}.

Here we transcribe the theory of the \YoungPOps\ into a form particularly well
suited to particle physics calculations, and show that the diagrammatic methods
of \refref{PCgr} can be profitably employed in explicit construction of \un\
multi-particle states, and evaluation of the associated \threenj\ coefficients.

In diagrammatic notation tensor objects are manipulated without any explicit
indices. Diagrammatic evaluation rules are intuitive and relations between
tensors can often be grasped visually. Take as an example the reduction of a
two-index tensor \(T_{ij}\) into symmetric and antisymmetric parts, \(T = (S +
A)T\), where
\begin{eqnarray*}
  ST_{ij} &=& 
    \frac12 \left(\id+(12)\right) T_{ij} \\
  AT_{ij} &=& \frac12 \left(\id-(12)\right) T_{ij} \,,
\end{eqnarray*}
and \(\id\) and \((12)\) denote the identity and the index transposition.
Diagrammatically, the two projection operators are drawn as
\begin{eqnarray}
  S &=& \frac12 \left\{ 
    \btrackYb{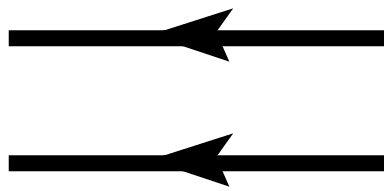}{6} + \btrackYb{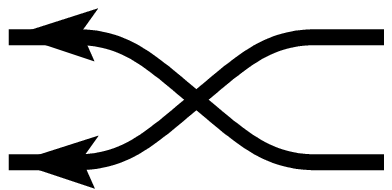}{6} \right\} \nonumber \\
  A &=& \frac12 \left\{ 
    \btrackYb{h2it.0}{6} - \btrackYb{h2it.1}{6}  \right\} \,.
  \label{birdie8.3}
\end{eqnarray}
It is clear at a glance that \(S\) symmetrizes and \(A\) antisymmetrizes the
two tensor indices. Here we shall construct such projection operator for
tensors of any rank.

R.~Penrose's papers are the first (known to the authors) to cast the
\YoungPOps\ into a diagrammatic form. Here we use Penrose diagrammatic notation
for symmetrization operators\rf{Penrose1971}, Levi-Civita
tensors\rf{PenroseMacCullen} and ``strand networks''\rf{Penrose71}. For several
specific, few-particle examples, diagrammatic \YoungPOps\ were constructed by
Canning\rf{Canning:1978ee}, Mandula\rf{Mandula} and Stedman\rf{StedmanBook}.
A diagrammatic construction of the
\un\ \YoungPOps\ for \emph{any} Young tableau
was outlined in the unpublished
\refref{ADKsl}, without proofs. Here we present the method in detail, as well
as the proof that the \YoungPOps\ so constructed are unique\rf{Elvang99}. The
other new results are the strand network derivation of the dimension formula
for \irrep s of \un, a proof that every \un\ \threenj\ coefficient is
proportional to the dimension of the largest \irrep\ within the \threenj\
diagram, and several sum rules for \un\ 3-\(j\) and 6-\(j\) coefficients.
 
The paper is organized as follows. The diagrammatic notation for tensors is
reviewed in \refsect{s-nota} and the Young tableaux in \refsect{s-YT}. This
material is standard and the reader is referred to any of the above cited
monographs for further details. In \refsect{YoungPOps} we construct
diagrammatic \YoungPOps\ for \un, and give formulas for the normalizations and
the dimensions of \un\ \irrep s. In \refsect{s:recoupling} we recast the
Clebsch--Gordan recoupling relations into a diagrammatic form, and show that
--- somewhat surprisingly --- the values of all \un\ \threenj\ coefficients
follow from the representation theory for the symmetric group
\(\symgrp_\numBoxes\) alone. The \threenj\ coefficients for \un\ are
constructed from the \YoungPOps\ and evaluated by diagrammatic methods in
\refsect{s-3j}. We derive a family of new sum rules for \un\ {\threenj}
coefficients in \refsect{s:sumrules}. In \refsect{s-adjoint} we briefly discuss
the case of \sun\ and mixed multi-particle anti-particle states. In
\refsect{s-negdim} we state and prove the negative dimensionality theorem for
\un. Not only does this proof provide an example of the power of diagrammatic
methods, but the theorem also simplifies certain group theoretic calculations.
We summarize our results in~\refsect{s:concl}.

The key, but lengthy original result presented in this paper, the proof of the
uniqueness, completeness, and orthogonality of the \YoungPOps\rf{Elvang99}, and
a strand network derivation of the dimension formula for \irrep s of \un\ are
relegated to \refappe{appendixA}.

\section{Diagrammatic notation} \label{s-nota}

In the diagrammatic notation\rf{PCgr} an invariant tensor is drawn as a
``blob'' with a leg representing each index. An arrow indicates whether it is
an upper or lower index; lower index arrows always point away from the blob
whereas upper index arrows point into the blob. The index legs are ordered in
the counterclockwise direction around the blob, and if the indices are not
cyclic there must be an indication of where to start, for example
\[T_{ab\;\;d}^{\;\;\;\,c\;\;e} = \btrackYb{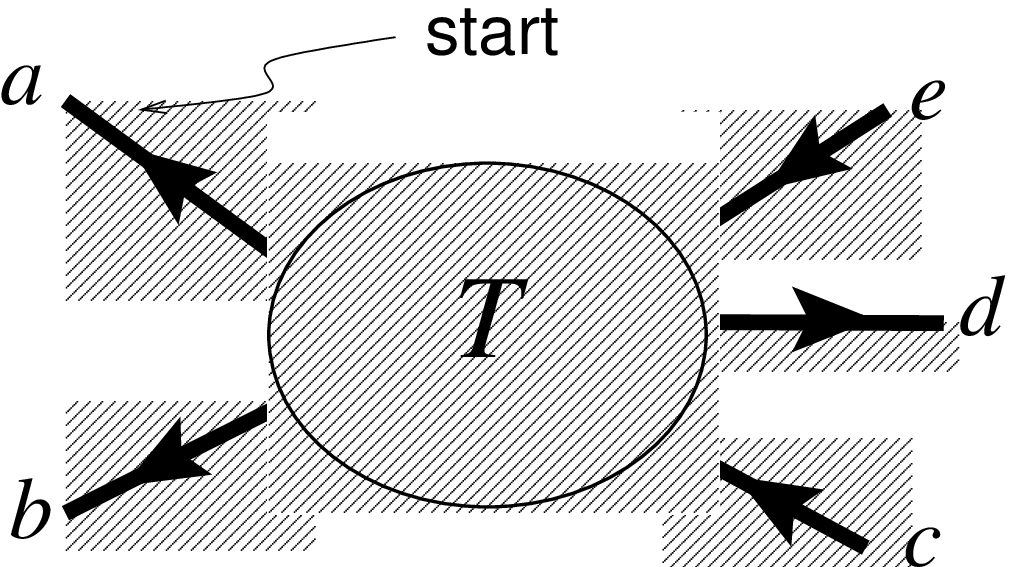}{16}\;.\] An internal line
in a diagram implies a sum over the corresponding index: matrix multiplication
is drawn as \[M_{a}^{\;\;b} N_{b}^{\;\;c} = \btrackYb{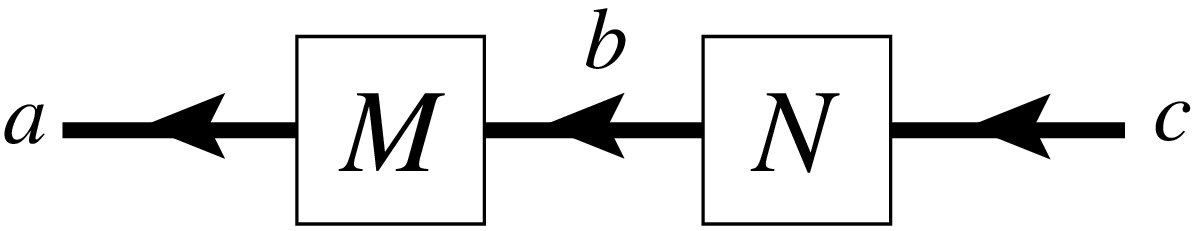}{19}\;,\] where
the index \(b\) can be omitted, as indeed can all other ``dummy'' indices. The
Kronecker delta is drawn as
\[
\delta_b^{a}
= b\btrackYb{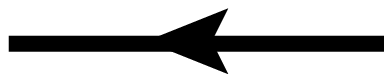}{6}a \quad a,b
= 1,2, \dots, n \,,\] and its trace --- the dimension of the representation ---
is drawn as a closed loop,
\begin{equation}
  \btrackYb{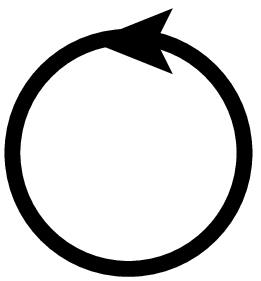}{4} = \delta_a^{a} =n \,.
  \label{loop}
\end{equation}

Index permutations can be drawn in terms of Kronecker deltas. For example, the
symmetric group \(\symgrp_2\) acting on two indices consists of the identity
element 
\(\id_{ab}^{\quad cd} = \delta_a^d \delta_b^c\)
and the
transposition \((12)_{ab}^{\quad cd} = \delta_a^c \delta_b^d\). In the
diagrammatic notation these operators are drawn as 
\[
\id_{ab}^{\quad cd}
= {a \atop b} \btrackYb{h2it.0}{6}{d \atop c} \quad \mbox{and} \quad
(12)_{ab}^{\quad cd} = {a \atop b}\btrackYb{h2it.1}{6}{d \atop c}\,.\]
Symmetrization of {\(p\)} indices is achieved by adding all permutations
\(\sigma\) of \(p\) indices, \(S=\frac1{p!} \sum_{\sigma\in \symgrp_p}
\delta_{\sigma(b_1}^{a_1} \!\cdots \delta_{b_p)}^{a_p}\). Similarly, the
operator \(A=\frac1{p!} \sum_{\sigma\in \symgrp_p} \sign{(\sigma)} \,
\delta_{\sigma(b_1}^{a_1}\!\cdots \delta_{b_p)}^{a_p}\) (with a minus for odd
permutations) antisymmetrizes \(p\) indices. Combinations of symmetrizers \(S\)
and antisymmetrizers \(A\) are collectively referred to as \emph{symmetry
operators}.

\begin{figure}[thb]
  \begin{tabular}{lcccllccc}
    (a) & \raisebox{0pt}[17pt][17pt]{\btrackYb{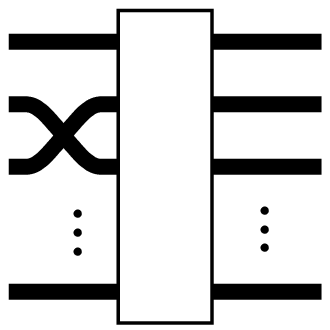}{5}} 
      & = & \btrackYb{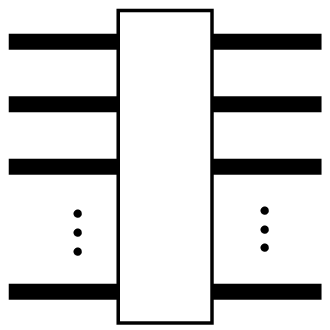}{5}   
      &\(\quad\)& (b) & \btrackYb{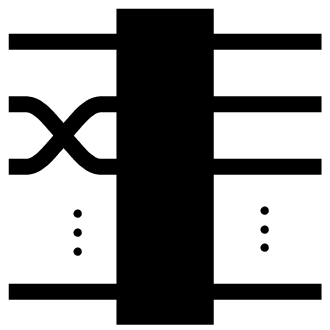}{5} 
      & = & \(-\)\btrackYb{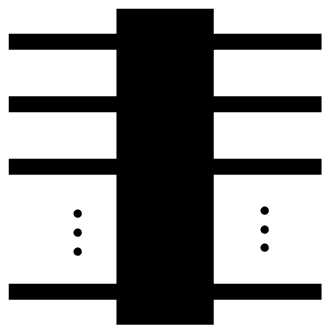}{5} \\
    (c) & \raisebox{0pt}[17pt][17pt]{\btrackYb{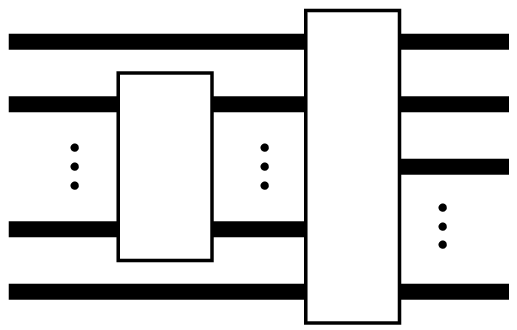}{8}} 
      & = & \btrackYb{hSprop.1b}{5}
      &\(\quad\)& (d) & \btrackYb{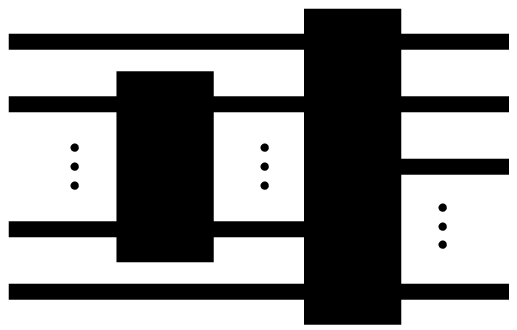}{8} 
      & = & \btrackYb{hAprop.1b}{5} \\
    (e) & \raisebox{0pt}[17pt][17pt]{\btrackYb{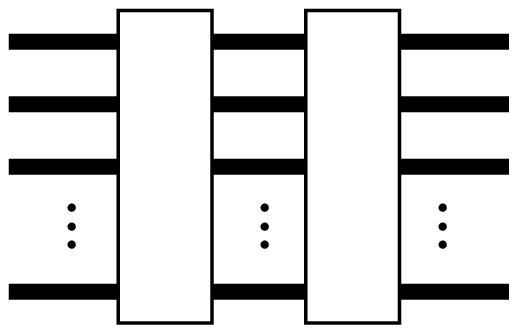}{8}} 
      & = & \btrackYb{hSprop.1b}{5}
      &\(\quad\)& (f) & \btrackYb{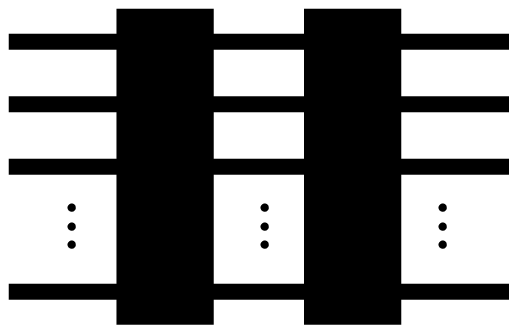}{8} 
      & = & \btrackYb{hAprop.1b}{5}  \\
  \end{tabular}
  \caption{Properties of the diagrammatic symmetrization and antisymmetrization
   operators.}
  \label{SA-prop}
\end{figure}

In the diagrammatic notation we write the symme\-trizers and the
antisymmetrizers of \emph{length} \(p\) as\rf{Penrose1971}
\begin{eqnarray} 
  \btrackYb{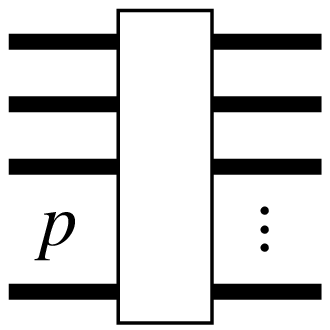}{5} & = & 
     \frac{1}{p!}\left( \btrackYb{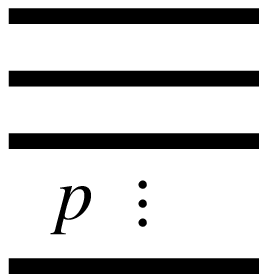}{4} 
     + \btrackYb{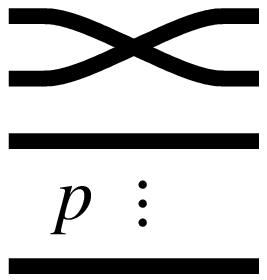}{4} + \cdots 
     + \btrackYb{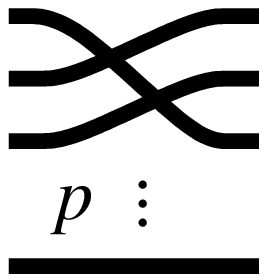}{4} + \cdots \right)
   \label{birdieS} \\
   \btrackYb{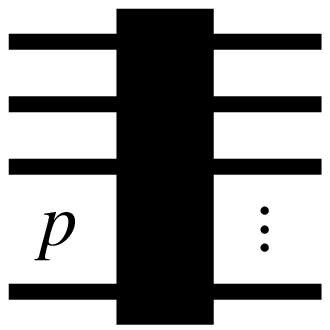}{5} & = & 
     \frac{1}{p!}\left( \btrackYb{hSn.Sa}{4} 
     - \btrackYb{hSn.Sb}{4} - \cdots 
     + \btrackYb{hSn.Sc}{4} + \cdots \right) \,.
   \label{birdieA}
\end{eqnarray}
In order to streamline the notation we shall neglect the arrows whenever this
leads to no confusion. Basic properties of the symmetry operators are listed in
\reffig{SA-prop}: A symmetrizer is invariant under any permutation of its legs,
rule~(a). The antisymmetrizer changes sign under odd permutations, rule~(b). A
symmetrizer connected by more than one line to an antisymmetrizer is zero by
rules~(a) and~(b),
\begin{equation}
  \btrackYb{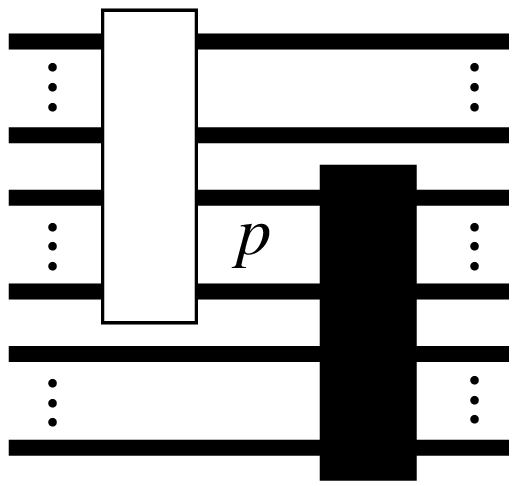}{8} = 0 \,, \quad p \ge 2 \,.
  \label{AS=0}
\end{equation}
Recursive identities for the (anti)symmetrizers are given in \refeq{smart-A}
and \refeq{smart-S}.

\section{Young tableaux} \label{s-YT}

Partition \(\numBoxes\) identical boxes into \(D\) subsets, and let
\(\lambda_m\), \(m=1,2,\dots,D\), be the number of boxes in the subsets ordered
so that \(\lambda_1 \ge \lambda_2 \ge \dots \ge \lambda_D \ge 1\). Then the
partition \(\lambda = [\lambda_1 , \lambda_2 , \dots , \lambda_D]\) fulfills
\(\sum_{m=1}^D \lambda_m = \numBoxes\). The diagram obtained by drawing the
\(D\) rows of boxes on top of each other, left aligned, starting with
\(\lambda_1\) at the top, is called a \emph{Young diagram}~\(\Y\).

Inserting each number from the set \(\{1,\dots,\numBoxes\}\) into a box of a
Young diagram \(\Y\) in such a way that numbers increase when reading a column
from top to bottom and numbers do not decrease when reading a row from left to
right yields a \emph{Young tableau} \(\Y_a\). The subscript \(a\) labels
different tableaux derived from a given Young diagram, i.e.,~different
admissible ways of inserting the numbers into the boxes. A \emph{standard
tableau} is a \(\numBoxes\)-box Young tableau constructed by inserting the
numbers \(1, \dots , \numBoxes\) according to the above rules, but using each
number exactly once.

As an example, three distinct standard tableaux, \[\btrackYb{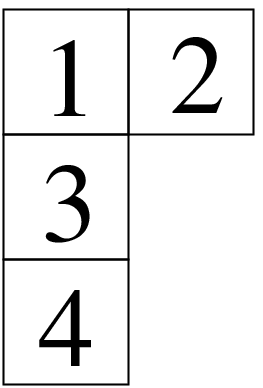}{4} \,,
\qquad \btrackYb{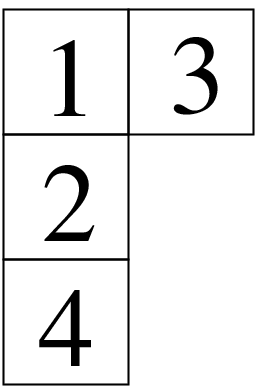}{4}\,, \qquad \btrackYb{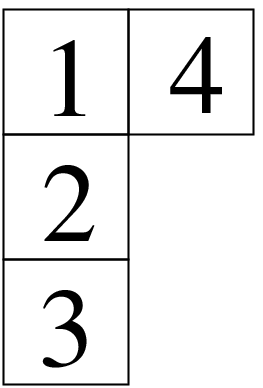}{4}\, ,\] are obtained
from the four-box Young diagram with partition \(\lambda = [2,1,1]\).

\subsection{Symmetric group \(\symgrp_\numBoxes\)}

Young diagrams label the \irrep s of the symmetric group \(\symgrp_\numBoxes\).
A \(\numBoxes\)-box Young diagram \(\Y\) corresponds to an \irrep\ of
\(\symgrp_\numBoxes\), and 
\(\Delta_\lambda\), the dimension of the \irrep\ \(\lambda\), is the number of
standard tableaux \(\Y_a\) that can be constructed from the Young diagram
\(\Y\). From the above example we see that the \irrep\ \(\lambda = [2,1,1]\) of
\(\symgrp_4\) is 3-dimensional. The formula for the dimension \(\Delta_\Y\) of
the \irrep\ of \(\symgrp_\numBoxes\) corresponding to the Young diagram \(\Y\)
is
\begin{equation}
  \Delta_\Y = \frac{k!}{|\Y|} \,.
  \label{Skdim}
\end{equation}
The number \(|\Y|\) is computed using a ``hook'' rule: Enter into each box of
the Young diagram the number of boxes below and to the left of the box,
including the box itself. Then \(|\Y|\) is the product of the numbers in all
the boxes. For instance, \[\Y = \btrackYb{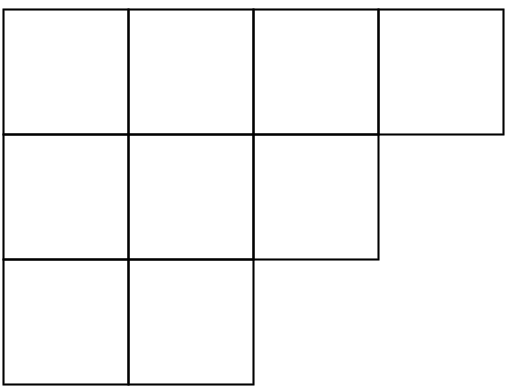}{8} ~~~~\longrightarrow~~~~
|\Y| = \btrackYb{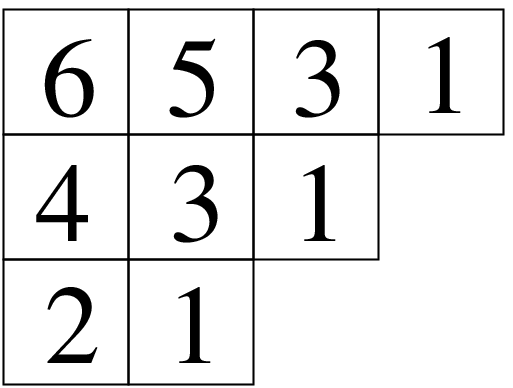}{8} = 6!\,3\,.\] The hook rule \refeq{Skdim} was first
proved surprisingly late, in 1954, by Frame, de B.~Robinson, and
Thrall\rf{Frame54}. Various proofs can be found in the
\refrefs{Robinson61,James81,Fulton91,Fulton99,GoodWallach,Bessen98}; in
particular, see Sagan\rf{Sagan01} and references therein.

\subsection{Representations of \un}

Whilst every Young diagram labels an \irrep\ of \(\symgrp_\numBoxes\), every
standard tableau labels an \irrep\ of \un. The dimension \(d_\Y\) of an
{\irrep} labeled by the Young diagram \(\Y\) equals the number of Young
tableaux \(\Y_a\) that can be obtained from \(\Y\) by inserting numbers from
the set \(\{1,2,\dots,n\}\) such that the numbers increase in each column and
do not decrease in each row.

For example, for \(SU(2)\) the partition \([2]\) corresponds to a 3-dimensional
\irrep\ with tableaux \btrackYt{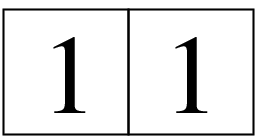}{2}, \btrackYt{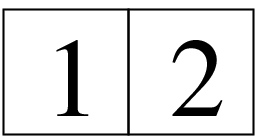}{2}, and
\btrackYt{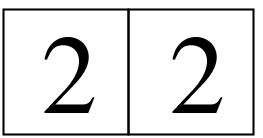}{2}, and the partition [1,1] corresponds to a 1-dimensional
\irrep\ with one tableau, \btrackYt{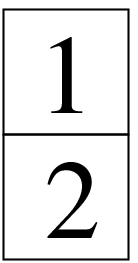}{1}. Similarly, one can check
that for \(SU(3)\), the partition \([2]\) is 6-dimensional and the partition
\([1,1]\) is 3-dimensional. We shall derive the dimension formula for any
{\irrep} of \un\ in \refsect{ss-prop}.

\section{Young projection operators} \label{YoungPOps}

We now present a diagrammatic method for construction of \YoungPOps. A
combinatorial version of these operators was given by van der
Waerden\rf{WAERDEN}, who credited von Neumann. There are many other versions in
the literature, all of them illustrating the fundamental theorem of 't~Hooft
and Veltman\rf{tHooftVelt73}: combinatorics cannot be taught. What follows
might aid those who think visually.

\subsection{The group algebra} \label{grpalg}

Our goal is to construct the projection operators such as \refeq{birdie8.3} for
any \irrep\ of \(\symgrp_\numBoxes\). We need to construct a basis set of
invariant tensors, multiply them by scalars, add and subtract them, and
multiply a tensor by another tensor. The necessary framework is provided by the
notion of {\em group algebra}.

The elements \(\sigma \in \symgrp_\numBoxes\) of the symmetric group
\(\symgrp_\numBoxes\) form a basis of a \(\numBoxes!\)-dimensional vector space
\(V\) of elements
\begin{equation}
  s = \sum_{\sigma \in \symgrp_\numBoxes} s_\sigma \, \sigma \, \in V \,,
   \label{eq:s}
\end{equation}
where \(s_\sigma\) are the components of the vector \(s\) in the given
basis. If \(s, t \in V\) have components \((s_\sigma)\) and \((t_\sigma)\), we 
define the product of \(s\) and \(t\) as the vector \(st\) in \(V\) with
components \((st)_\sigma = \sum_{\tau \, \in \symgrp_\numBoxes} s_\tau\,
t_{\tau^{-1}\, \sigma}\). This multiplication is associative because it relies
on the associative group operation. Since \(V\) is closed under the
multiplication the elements of \(V\) form an associative algebra --- the
\emph{group algebra} of \(\symgrp_\numBoxes\). Acting on an element \(s \in V\)
with any group element maps \(s\) to another element in the algebra, hence this
map gives a \(\numBoxes!\)-dimensional matrix representation of the group
algebra, the {\em regular representation}. Note that the matrices of any
representation \(\mu\) of the group is also a basis for the representation of
the algebra: Let \(D^\mu(\sigma)\) denote a (possibly reducible) representation
of \(\symgrp_\numBoxes\). The group algebra of \(\symgrp_\numBoxes\) in the 
representation \(\mu\) then consists of elements \[D^\mu(s) = \sum_{\sigma \in
\symgrp_\numBoxes} s_\sigma \, D^\mu(\sigma)\, \in V \,,\] where \(s\) is given
by \refeq{eq:s}. 
The minimal left-ideals \(V_\lambda\) of the group algebra (i.e., \(s V_\lambda
= V_\lambda\) for all \(s \in V\), and \(V_\lambda\) has no proper subideals)
are the proper invariant subspaces corresponding to the \irrep s of the
symmetric group \(\symgrp_\numBoxes\).

The regular representation is reducible and each {\irrep} appears
\(\Delta_\lambda\) times in the reduction, where \(\Delta_\lambda\) is the
dimension of the subspace \(V_\lambda\) corresponding to the \irrep\
\(\lambda\). This gives the well-known relation between the order of the
symmetric group \(|\symgrp_\numBoxes|=\numBoxes!\) (the dimension of the
regular representation) and the dimensions of the \irrep s,
\[|\symgrp_\numBoxes| = \sum_{\rom{irreps}~\lambda} \Delta_\lambda^2 \,.\]
Using \refeq{Skdim} and the fact that the Young diagrams label the \irrep s of
\(\symgrp_\numBoxes\), we have
\begin{equation}
  1 = k! \sum_{(k)} \frac{1}{|\Y|^2} \,,
  \label{addition}
\end{equation}
where the sum is over all Young diagrams with \(\numBoxes\) boxes. We shall use
this relation to determine the normalization of \YoungPOp s in
\refappe{appendixA}.

The reduction of the regular representation of \(\symgrp_\numBoxes\) gives a
completeness relation \[\id = \sum_{(k)} P_\Y\] into projection operators
\[P_\Y = \sum_{\Y_a \in \Y} P_{\Y_a} \,.\] The sum is over all Young tableaux
derived from the Young diagram \(\Y\). Each \(P_{\Y_a}\) projects onto the
corresponding invariant subspace \(V_{\Y_a}\) --- for each \(\Y\) there are
\(\Delta_\Y\) such projection operators (corresponding to the \(\Delta_\Y\)
possible standard arrangements of the diagram) and each of these project onto
one of the \(\Delta_\Y\) invariant subspaces \(V_\Y\) of the reduction of the
regular representation. It follows that the projection operators are orthogonal
and that they constitute a complete set.

\subsection{Diagrammatic \YoungPOp s} \label{s-constr}

We now generalize \refeq{birdie8.3}, the \(\symgrp_2\) projection operators
expressed in terms of Kronecker deltas, to \YoungPOp\ for
any~\(\symgrp_\numBoxes\).

The Kronecker delta is invariant under unitary transformations, \(\delta_a^b =
(U^{\dagger}){}_a{}^{a'} \delta^{b'}_{a'} U{}_{b'}{}^b\), \(U\in U(n)\), and so
is any combination of Kronecker deltas, such as the symmetrizers of
\reffig{SA-prop}. Since these operators constitute a complete set, any \un\
invariant tensor built from Kronecker deltas can be written in terms of
symmetrizers and antisymmetrizers. In particular, the invariance of the
Kronecker delta under {\un} transformations implies that the same symmetry
group operators which project the \irrep s of \(\symgrp_\numBoxes\) also yield
the \irrep s of \un.

The simplest examples of \YoungPOps\ are those associated with the Young
tableaux consisting of either one row or one column. The corresponding
\YoungPOps\ are simply the symmetrizers \refeq{birdieS} or the antisymmetrizers
\refeq{birdieA}, respectively. As projection operators for
\(\symgrp_\numBoxes\), the symmetrizer projects onto the one dimensional
subspace corresponding to the fully symmetric representation, and the
antisymmetrizer projects onto the alternating representation.

A \YoungPOp\ for a mixed symmetry Young tableau will here be constructed by
first antisymmetrizing subsets of indices, and then symmetrizing other subsets
of indices; which subsets is dictated by the form of the Young tableau, as will
be explained shortly. Schematically, \[P_{\Y_a} = \alpha_\Y
\btrackYb{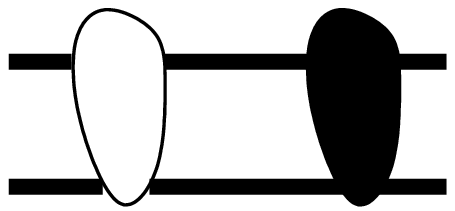}{7}\,,\] where \(\alpha_\Y\) is a normalization constant
(defined below) ensuring that the operators are idempotent, \(P_{\Y_a} P_{\Y_b}
= \delta_{ab} P_{\Y_a}\). This particular form of projection operators is by no
means unique --- \YoungPOp\ symmetric under transposition are constructed in
\refref{PCgr} --- but is particularly convenient for explicit computations.

Let \(\Y_a\) be a \(\numBoxes\)-box standard tableau. Arrange a set of
symmetrizers corresponding to the rows in \(\Y_a\), and to the right of this
arrange a set of antisymmetrizers corresponding to the columns in \(\Y_a\). For
a Young diagram \(\Y\) with \(s\) rows and \(t\) columns we label the rows
\(\row_1\), \(\row_2\), \dots , \(\row_s\) and to the columns \(\col_1\),
\(\col_2\), \dots, \(\col_t\). 
Each symmetry operator in \(P_{\Y_a}\) is associated to a row/column in
\(\Y_a\), hence we label a symmetry operator after the corresponding
row/column, for example \[\btrackYb{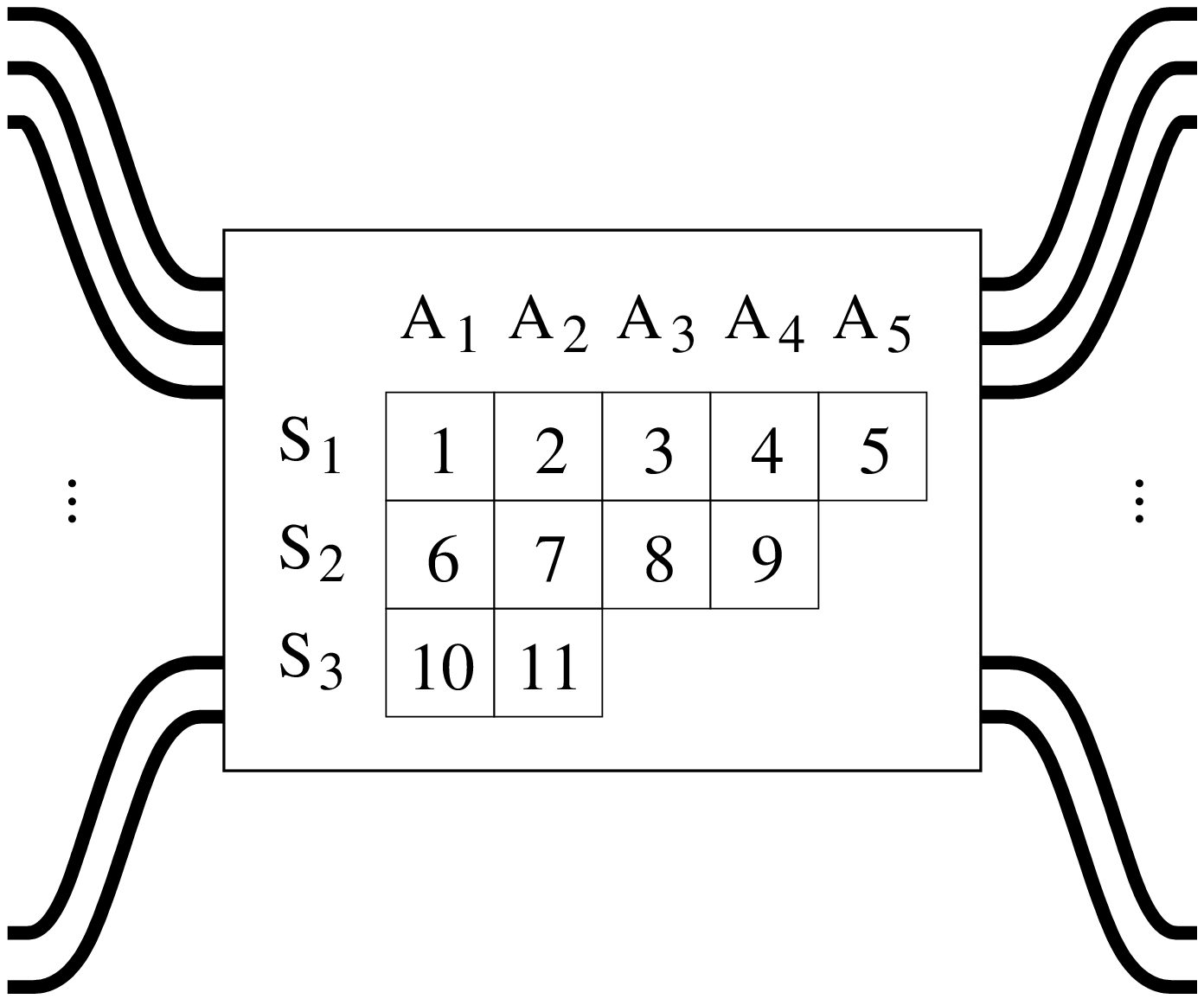}{22} \;=\; \alpha_\Y
\btrackYb{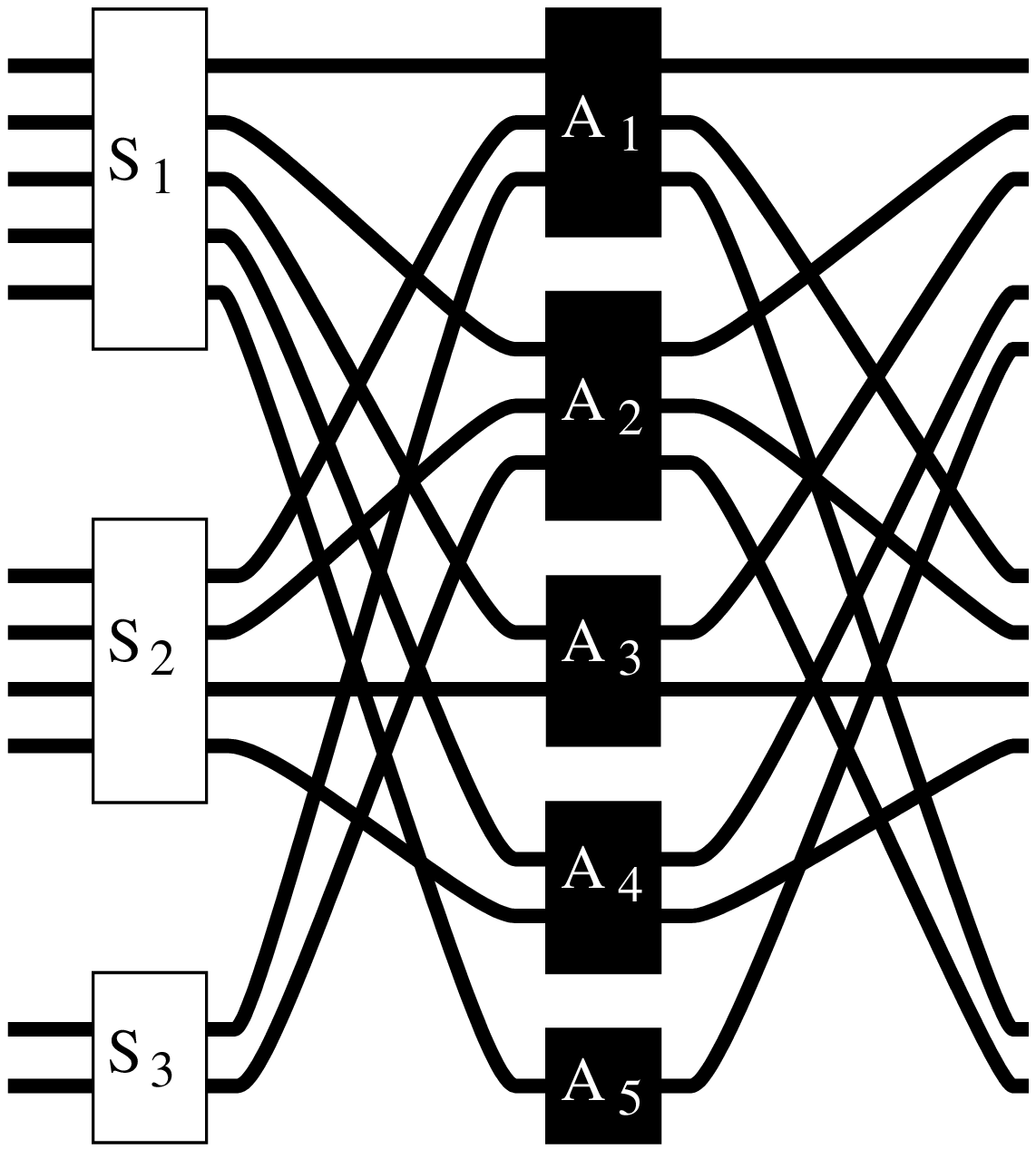}{18}\,.\] Let the lines numbered 1 to \(\numBoxes\) enter the
symmetrizers as described by the numbers in the boxes in the standard tableau
and connect the set of symmetrizers to the set of antisymmetrizers in a
non-vanishing way, avoiding multiple intermediate lines prohibited by
\refeq{AS=0}. Finally, arrange the lines coming out of the antisymmetrizers
such that if the lines all passed straight through the symmetry operators, they
would exit in the same order as they entered.

We shall denote by \(\Delta_\Y\) the dimensions of \irrep s of
\(\symgrp_\numBoxes\), and by \(d_\Y\) the dimensions of \irrep s of \un. Let
\(|\row_i|\) or \(|\col_i|\) denote the number of boxes within a row or column,
respectively. Thus \(|\col_i|\) also denotes the number of lines entering the
antisymmetrizer \(A_i\), and similarly for the symmetrizers. The normalization
constant \(\alpha_\Y\) is given by \[\alpha_\Y = \frac{\prod_{i=1}^s |\row_i|!
\prod_{j=1}^t |\col_j|!}{|\Y|} \,,\] where \(|\Y|\) is related through
\refeq{Skdim} to \(\Delta_\Y\), the dimension of \irrep\ \(\Y\) of
\(\symgrp_\numBoxes\), and is a hook rule \(\symgrp_\numBoxes\) combinatoric
number. The normalization depends only on the shape of the Young diagram, not
the particular tableau. The \YoungPOps
\renewcommand{\labelenumi}{\theenumi)}
\begin{enumerate}
\item are \emph{idempotent}, \(P_\Y^2 = P_\Y\)
\item are \emph{orthogonal}: If \(\Y\) and \(\Z\) are two distinct standard
  tableaux, then \(P_\Y P_\Z = P_\Z P_\Y = 0\) , and
\item constitute a \emph{complete set}, \(\id = \sum P_\Y\), where
  the sum is over all standard tableaux \(\Y\) with \(\numBoxes\) boxes.
\end{enumerate}

The projections are unique up to an overall sign. By construction, the identity
element always appears as a term in the expansion of the symmetry operators of
the \YoungPOps\ --- the overall sign is fixed by requiring that the identity
element comes with a positive coefficient. The diagrammatic proof that the
above rules indeed assign a unique projection operator to each standard
tableaux is the central result of this paper; as it would impede the flow of
our argument at this point, it is placed into \refappe{a_unique}.

\noindent {\bf Example}: The Young diagram corresponding to the partition
\([3,1]\) tells us to use one symmetrizer of length three, one of length one,
one antisymmetrizer of length two, and two of length one. There are three
distinct standard tableaux, each corresponding to a projection operator
\begin{eqnarray*}
  \btrackYb{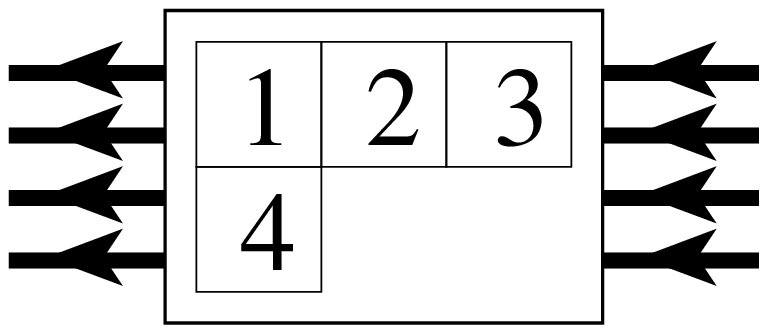}{12} & = & \alpha_\Y \, \btrackYb{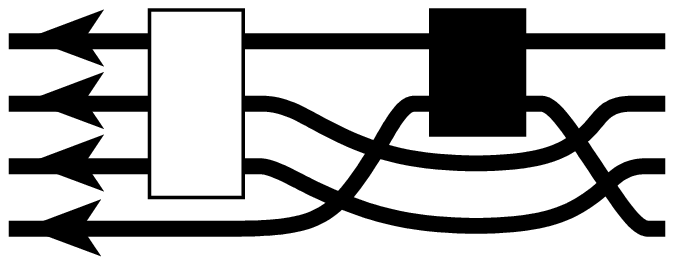}{11} \\
  \btrackYb{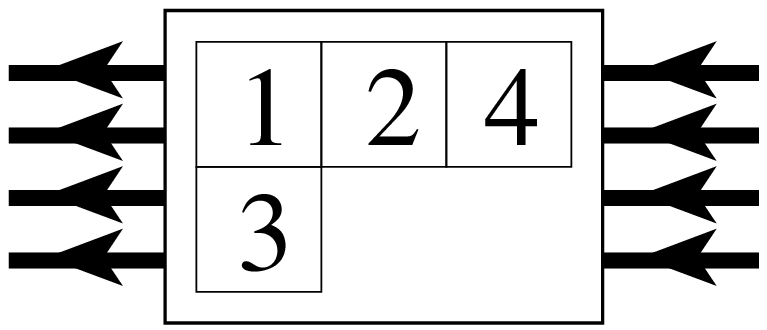}{12} & = & \alpha_\Y \, \btrackYb{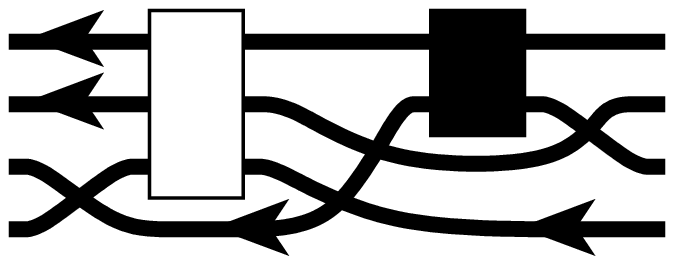}{11} \\
  \btrackYb{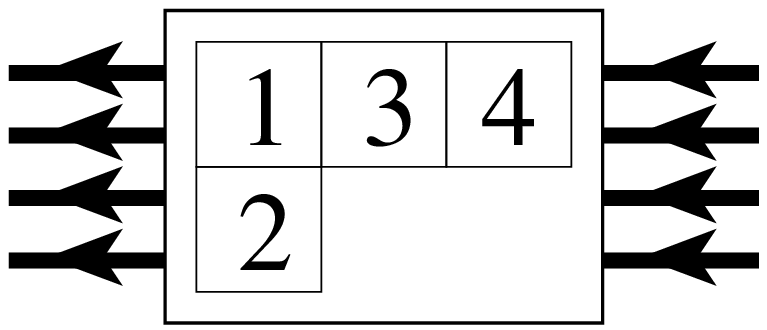}{12} & = & \alpha_\Y \, \btrackYb{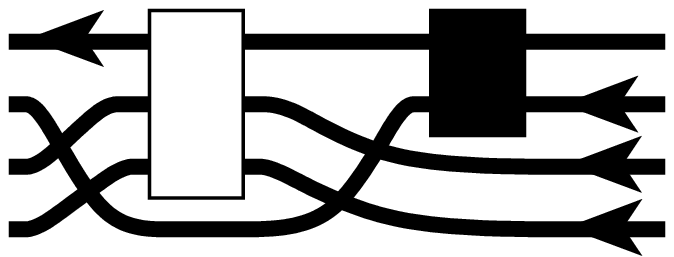}{11} \;,
\end{eqnarray*}
where \(\alpha_\Y\) is a normalization constant. The symmetry operators of
unit width need not be drawn explicitly. We have \(|\Y| = 8\),
\(|\row_1| = 3\), \(|\row_2| = 1\), \(| \col_1| = 2\), etc, yielding
the normalization \(\alpha_\Y = 3/2\).

\subsection{Dimensions of \un\ \irrep s} \label{ss-prop}

The dimension \(d_\Y\) of a \un\ \irrep\ is computed by taking the trace of the
corresponding \YoungPOp, \(d_\Y = \tr P_\Y\). The trace can be evaluated by
expanding the symmetry operators using \refeq{birdieS} and \refeq{birdieA}. By
\refeq{loop}, each closed line is worth \(n\), so \(d_\Y\) is a polynomial in
\(n\) of degree \(\numBoxes\).

\noindent {\bf Example}: The dimension of a three-index \YoungPOp:
\begin{eqnarray}
   d_\Y & = & \btrackYb{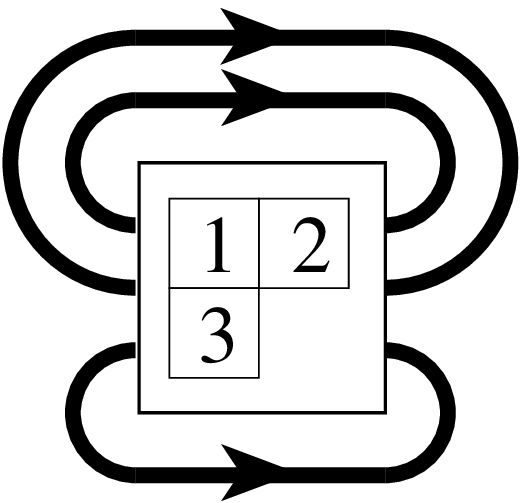}{8} = \frac43\, \btrackYb{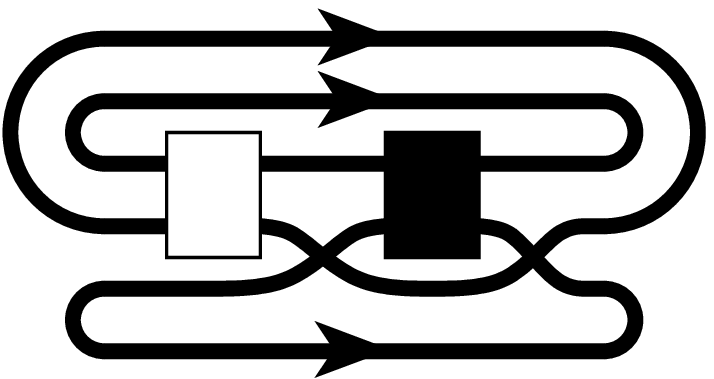}{11} 
     \nonumber \\
     & = & \frac43\left( \frac1{2!}\right)^2
       \left( \btrackYb{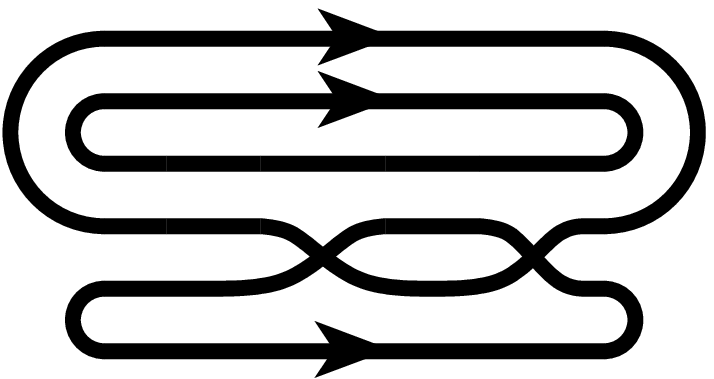}{11} + \btrackYb{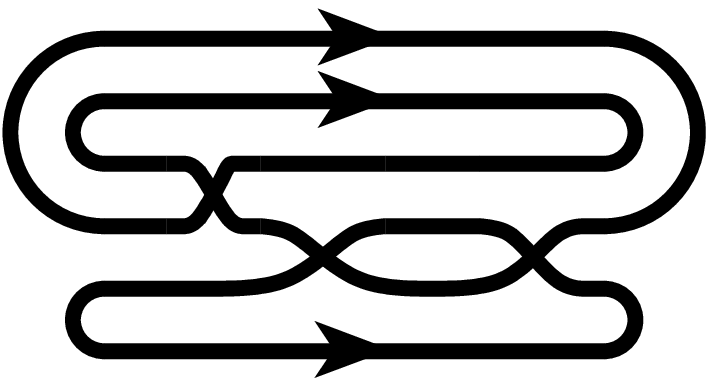}{11} \right. \\
   && \left. - \btrackYb{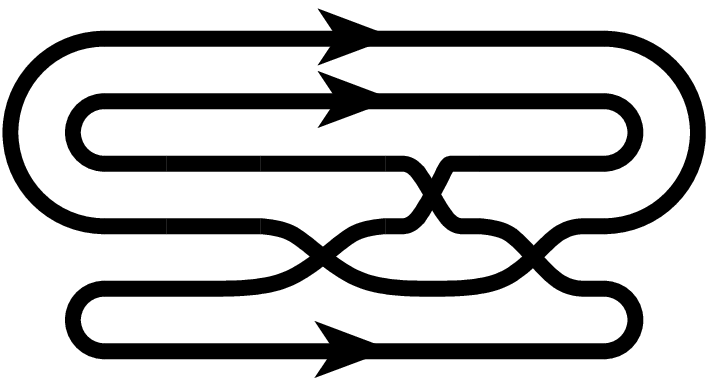}{11} - \btrackYb{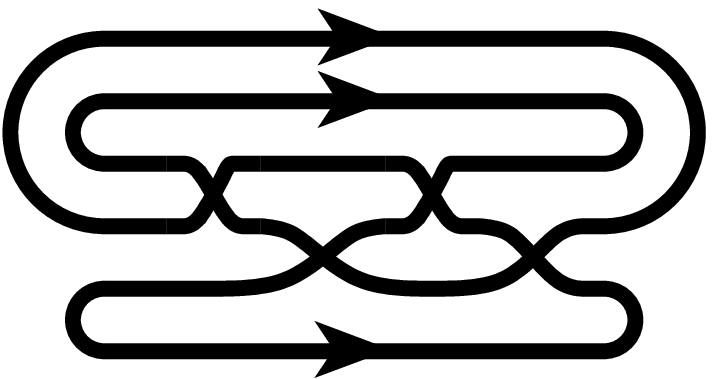}{11} \right)
   \label{expand} \\ 
   & = & \frac{1}{3} (n^3 + n^2 - n^2 -n) = \frac{n(n^2 -1)}{3} \,.
   \nonumber
\end{eqnarray}
Such brute expansion is unnecessarily laborious: The dimension of the \irrep\
labeled by \(\Y\) is
\begin{eqnarray}
  d_\Y = \frac{f_\Y(n)}{|\Y|} \,,
  \label{dimenUn}
\end{eqnarray}
where \(f_\Y(n)\) is the polynomial in \(n\) obtained from the Young diagram
\(\Y\) by multiplying the numbers written in the boxes of \(\Y\), according to
the following rules: (A) The upper left box contains an \(n\). (B) The numbers
in a row increase by one when reading from left to right. (C) The numbers in a
column decrease by one when reading from top to bottom. Hence, if \(\numBoxes\)
is the number of boxes in \(\Y\), \(f_\Y(n)\) is a polynomial in \(n\) of
degree \(\numBoxes\). The dimension formula \refeq{dimenUn} is well-known, see
for instance \refref{Georgi99}.

In the example \refeq{expand}, we have \(f_\Y(n) = n(n-1)(n+1)\) and \(|\Y| =
3\), giving \(d_\Y = \frac{n(n^2 -1)}{3}\).

\noindent {\bf Example}: For \(\Y = [4,2,1]\) we have \[d_\Y =
\frac{\btrackYb{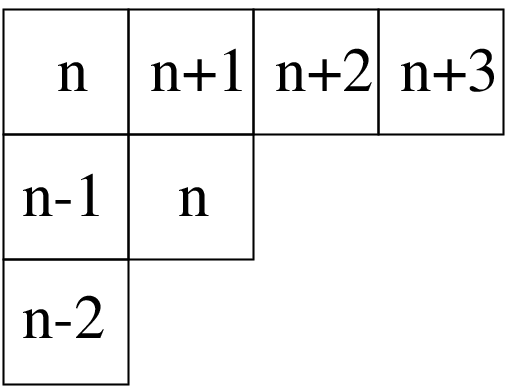}{8}}{\btrackYb{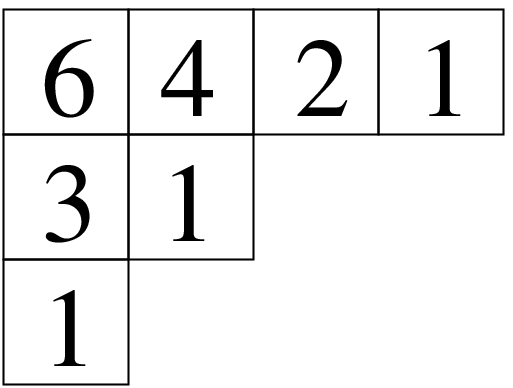}{8}} = \frac{n^2(n^2-1)
(n^2-4) (n+3)}{144}\,.\] A diagrammatic proof of the \un\ dimension formula
\refeq{dimenUn} is given in \refappe{a_dim}.

Diagrammatically, the number \(f_\Y(n)\) is the number of \(n\)-color colorings
of the strand network corresponding to \(\tr P_\Y\). Each strand is a closed
path passing straight through each symmetry operator. The number of strands
equals \(\numBoxes\), the number of boxes in \(\Y\). The top strand
(corresponding to the leftmost box in the first row of \(\Y\)) may be colored
in \(n\) ways. Color the rest of the strands according to
\begin{description}
\item[Rule 1:] If a path which could be colored in \(m\) ways enters an
  antisymmetrizer, the lines below it can be colored in \(m-1, m-2, \dots\)
  ways.
\item[Rule 2:] If a path which could be colored in \(m\) ways enters a
  symmetrizer, the lines below it can be colored in \(m+1, m+2, \dots\) ways.
\end{description}
The number of ways to color the strand diagram is \(f_\Y(n)\) as defined above.

\noindent {\bf Example}: For \(\Y\) = \btrackYb{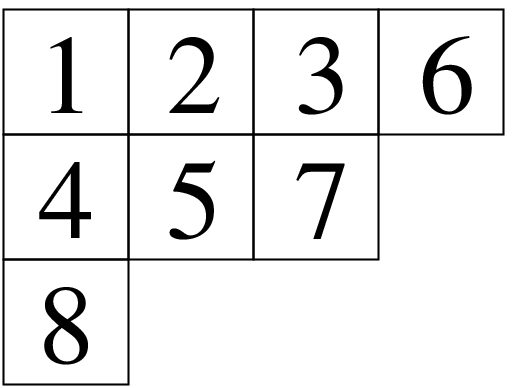}{6}, the strand
diagram is \[\btrackYb{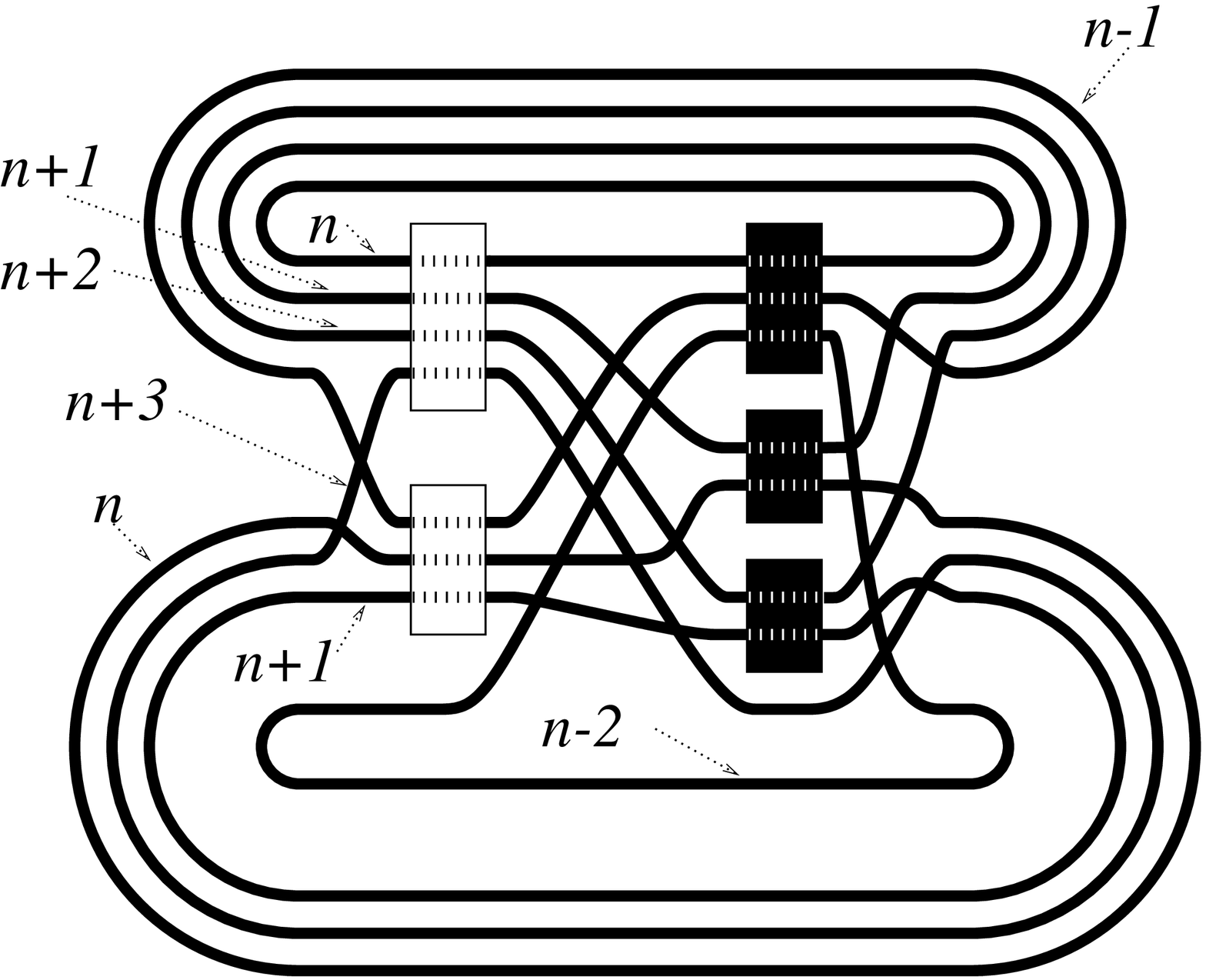}{32}\,.\] Each strand is labeled by the
number of admissible colorings. Multiplying these numbers and including the
factor \(1/|\Y|\), we find
\begin{eqnarray*}
  d_\Y &=& \frac{(n-2)\, (n-1)\, n^2\, (n+1)^2\, (n+2)\, (n+3)}
    {\btrackYb{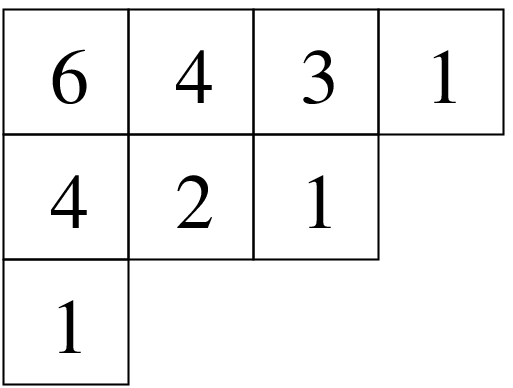}{6}} \\
  &=& \frac{n\, (n + 1)\, (n+3)!}{2^6 \, 3^2 \, (n-3)! }
\end{eqnarray*}
in agreement with \refeq{dimenUn}.

\subsection{Examples} \label{s-example}

We present examples to illustrate decomposition of reducible representation
into \irrep s (plethysm) using the diagrammatic projection operators.

The Young diagram \btrackYb{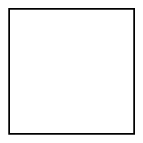}{2} corresponds to the fundamental
\(n\)-dimensional \irrep\ of \un. As we saw in \refeq{birdie8.3}, the direct
product of two of these \(n\)-dimensional representations is a
\(n^2\)-dimensional reducible representation,
\begin{eqnarray}  
  \btrackYb{Ybox}{2} \otimes \btrackYb{Ybox}{2}
    &=& \hspace{3.3mm} \btrackYb{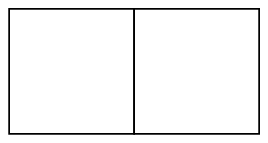}{4} \hspace{3.5mm} \oplus 
    \hspace{3.5mm} \btrackYb{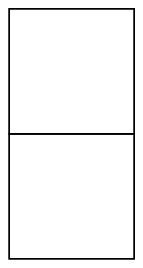}{2} 
  \label{2linesplita} \\ 
  \btrackYb{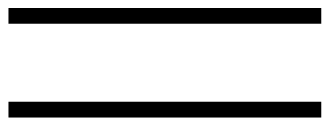}{5} \hspace{1.5mm} 
    &=& \hspace{2.5mm} \btrackYb{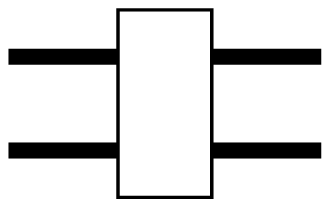}{5} \hspace{2.6mm} + 
      \hspace{2.5mm} \btrackYb{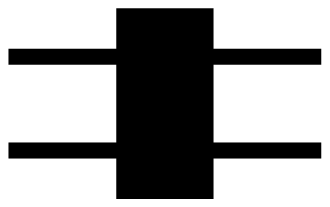}{5}  
  \label{2linesplitc} \\ 
  n^2 \quad &=& \frac{n(n+1)}{2} + \frac{n(n-1)}{2} \,.
  \label{2linesplitb}
\end{eqnarray}
Eq.~\refeq{2linesplita} shows the decomposition of the reducible representation
in terms of Young diagrams, and \refeq{2linesplitc} gives the corresponding
projection operators. Tracing \refeq{2linesplitc} yields the dimensions
\refeq{2linesplitb} of the \irrep s.

\begin{figure}[bth] 
  \begin{center}
    \begin{tabular}{|c|c|c|} 
      \hline 
      \(\Y_a\) & \(d_{\Y_a}\) 
        & \raisebox{0pt}[10pt][5pt]{\(P_{\Y_a}\)} \\
      \hline
      \btrackYb{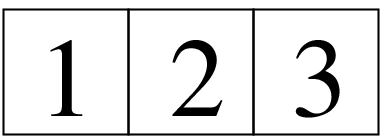}{6} & \(\frac{n(n+1)(n+2)}{6}\) 
        & \raisebox{0pt}[15pt][15pt]{\btrackYb{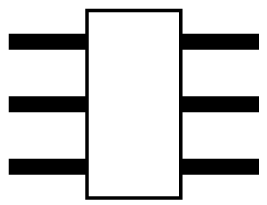}{4}} \\
      \btrackYb{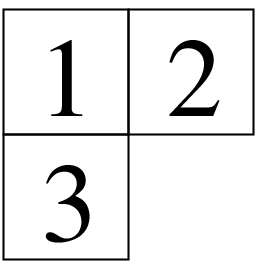}{4} & \(\frac{n(n^2-1)}{3}\)    
        & \(\frac{4}{3}\)\raisebox{0pt}[15pt][15pt]{\btrackYb{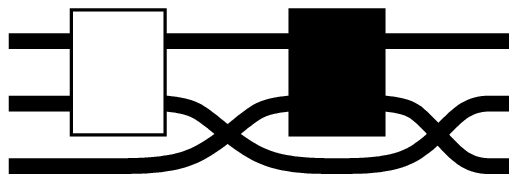}{8}} \\ 
      \btrackYb{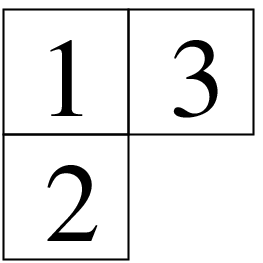}{4} & \(\frac{n(n^2-1)}{3}\)
        & \(\frac{4}{3}\)\raisebox{0pt}[15pt][15pt]{\btrackYb{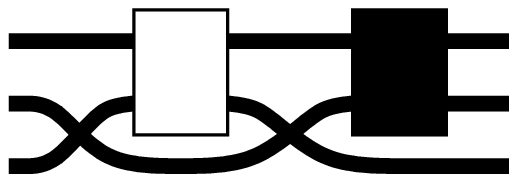}{8}} \\ 
      \btrackYb{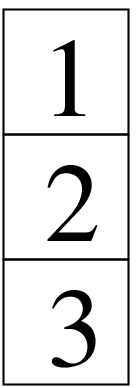}{2} & \(\frac{(n-2)(n-1)n}{6}\)
        & \raisebox{0pt}[15pt][15pt]{\btrackYb{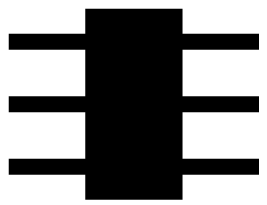}{4}} \\ 
      \hline 
      \btrackYb{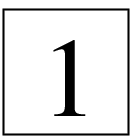}{2} \(\otimes\) \btrackYb{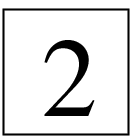}{2} 
        \(\otimes\) \btrackYb{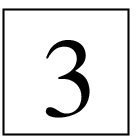}{2} & \(n^3\) 
        & \raisebox{0pt}[15pt][9pt]{\btrackYb{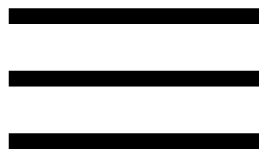}{4}} \\
      \hline
   \end{tabular}
   \vspace{0.2cm}
   \caption{Reduction of a three-index tensor. Bottom row: the direct product
     of three unit tableaux, the sum of dimensions, and the the projection
     operators completeness sum.}
   \label{red3it}
 \end{center}
\end{figure}

The first non-trivial example is the reduction of the three-index tensor
\YoungPOps, listed in \reffig{red3it}. Further examples can be found in
\refref{PCgr}.

The four projectors are orthogonal by inspection. In order to verify the
completeness, expand first the two three-index projection operators of mixed
symmetry: \[\frac43 \left( \btrackYb{h3it.nP2}{8} + \btrackYb{h3it.nP3}{8}
\right)\]
\begin{equation}
  \qquad\quad = \frac{2}{3}  \btrackYb{h3it.nid}{4} 
    - \frac13\btrackYb{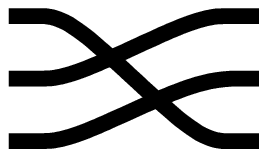}{4} 
    - \frac13\btrackYb{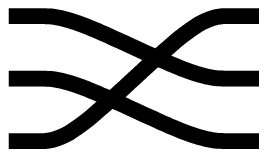}{4} \, .
  \label{ex.compl.b} 
\end{equation}
In the sum of the fully symme\-tric and the fully antisymmetric tensors all the
odd permutations cancel, and we are left with
\begin{equation}
  \btrackYb{h3it.nP1}{4} + \btrackYb{h3it.nP4}{4} 
    = \frac{1}{3} \left( \btrackYb{h3it.nid}{4} + \btrackYb{hperm.123}{4} 
      + \btrackYb{hperm.132}{4} \right)\,.
  \label{ex.compl.a}
\end{equation}
Adding \refeq{ex.compl.b} and \refeq{ex.compl.a} we find \[\btrackYb{h3it.nP1}
{4} + \frac{4}{3}\btrackYb{h3it.nP2}{8} + \frac{4}{3}\btrackYb{h3it.nP3}{8} +
\btrackYb{h3it.nP4}{4} = \btrackYb{h3it.nid}{4} \,,\] verifying the
completeness relation.

Acting with any permutation on the fully sym\-metric or antisymmetric
projection operators gives \(\pm 1\) times the projection operator (see
\reffig{SA-prop}). For projection operators of mixed symmetry the action of a
permutation is not as simple, because the permutations will mix the spaces
corresponding to the different tableaux. Here we shall need only the action of
a permutation within a {\threenj} coefficient, and, as we shall show below, in
this case the result will again be simple, a factor \(\pm 1\) or~\(0\).

\section{Recoupling relations} \label{s:recoupling} 

In the spirit of Feynman diagrams, group theoretic weights with all indices
contracted can be drawn as ``vacuum bubbles''. We now show that for \un\ any
such vacuum bubble can be evaluated diagrammatically, either directly, as a
\threenj\ coefficient, or following a reduction to 3-\(j\) and 6-\(j\)
coefficients. The exposition of this section follows closely \refref{PCgr}; the
reader can find there more details, as well as the precise relationship between
our 3-\(j\) and 6-\(j\) coefficients, and the Wigner 3-\(j\) and 6-\(j\)
symbols\rf{WIGNER}.

The decomposition of a many-particle state can be implemented sequentially,
decomposing two-particle states at each step. The Clebsch-Gordan coefficients
for \(\X\otimes\Z\to\Y\) can be drawn as {\em 3-vertices}
\begin{equation}
  {\frac1{\sqrt{a}} }\; {\btrackYb{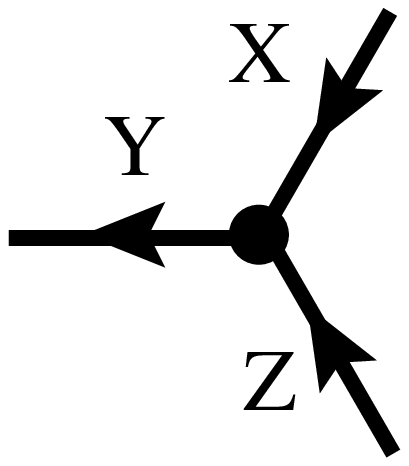}{6}} \,,
  \label{3vertex}
\end{equation}
where \({1}/{\sqrt{a}}\) is an (arbitrary) normalization constant. The
projection operators for \(\X \otimes \Z \to \Y\to\X\otimes\Z\) can
be drawn as \[\frac1a \; \btrackYb{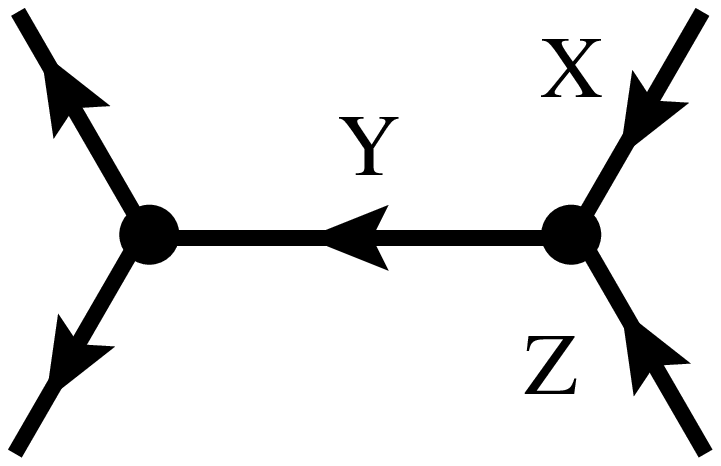}{11}\,.\] The orthogonality of
\irrep s implies \(\W=\Y\) in
\begin{equation}
  \btrackYb{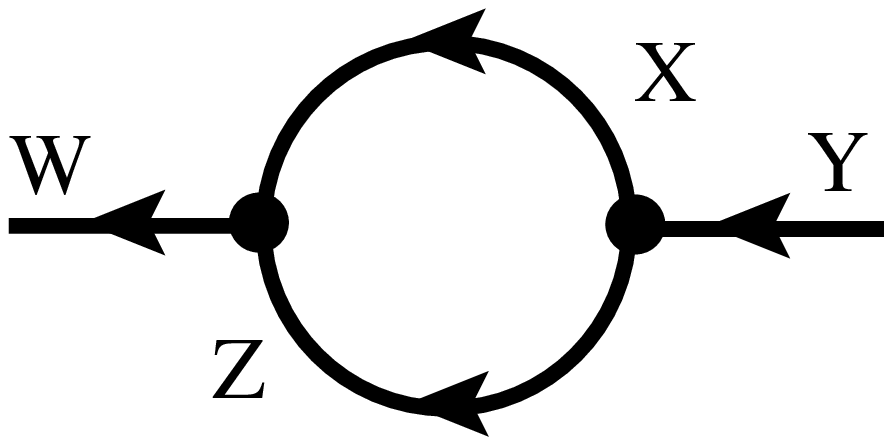}{14} = a \; \btrackYb{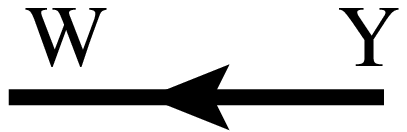}{6} \;,
  \label{B3vort}
\end{equation}
and the completeness relation can be drawn as
\begin{equation}
  \btrackYb{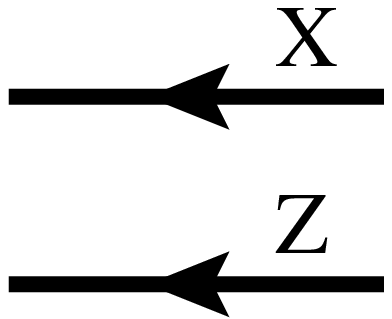}{6} 
    = \sum_{\Y} \frac1{a_\Y} \; \btrackYb{hre.proj}{11} \,,
  \label{Bprojcompl}
\end{equation}
where the sum is over all \irrep s contained in \(\X \otimes \Z\).

The normalization constant \(a\) can be computed by tracing \refeq{B3vort},
\[\btrackYb{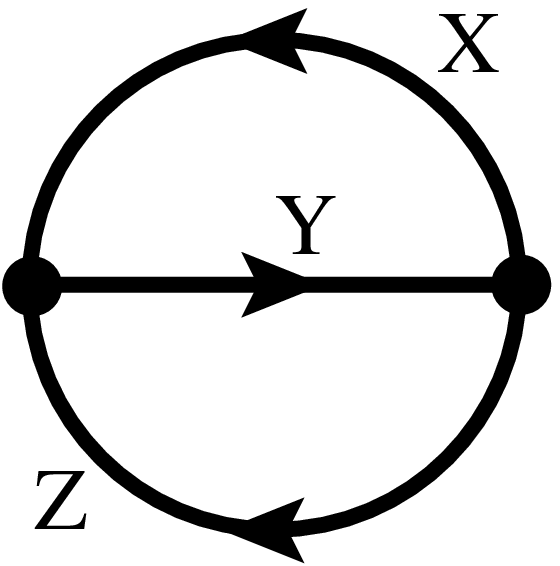}{7} =  a\;\btrackYb{hloop}{4} = ad_\Y\;,\] where
\(d_\Y\) is the dimension of the representation \(\Y\). The vacuum
bubble on the left hand side is called a 3-\(j\) coefficient. More generally,
vacuum bubbles with \(n\) lines are called \threenj\ coefficients.
 
\begin{figure}[thb]
  \begin{center}
    \btrackYb{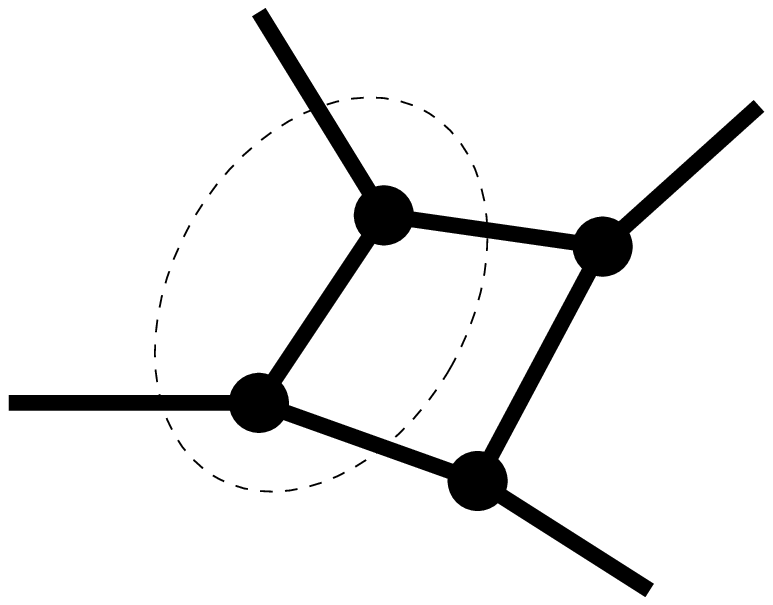}{10} \(\to\) \btrackYb{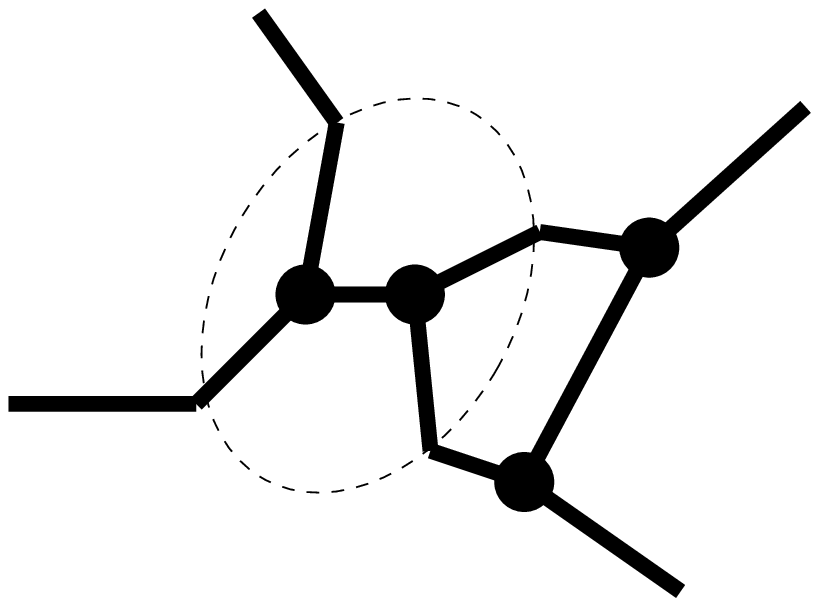}{11} 
    \(\to\) \btrackYb{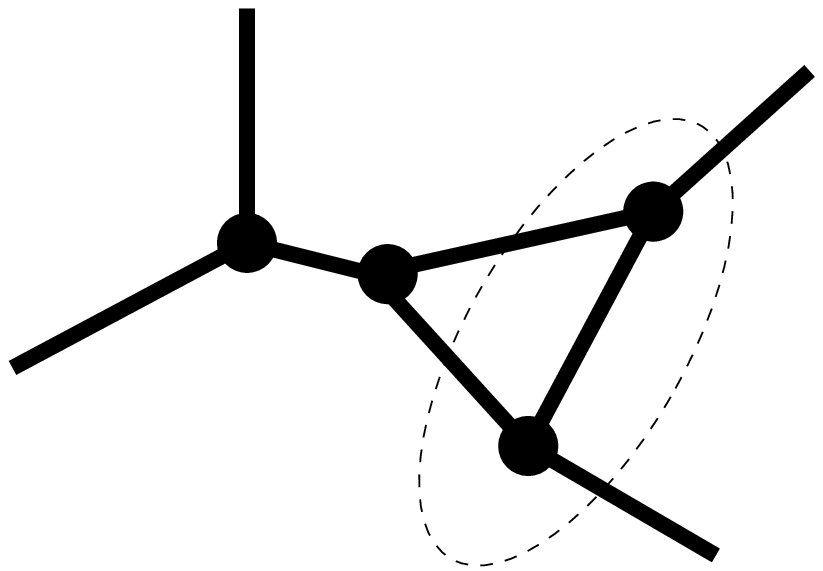}{11} \(\to\) \btrackYb{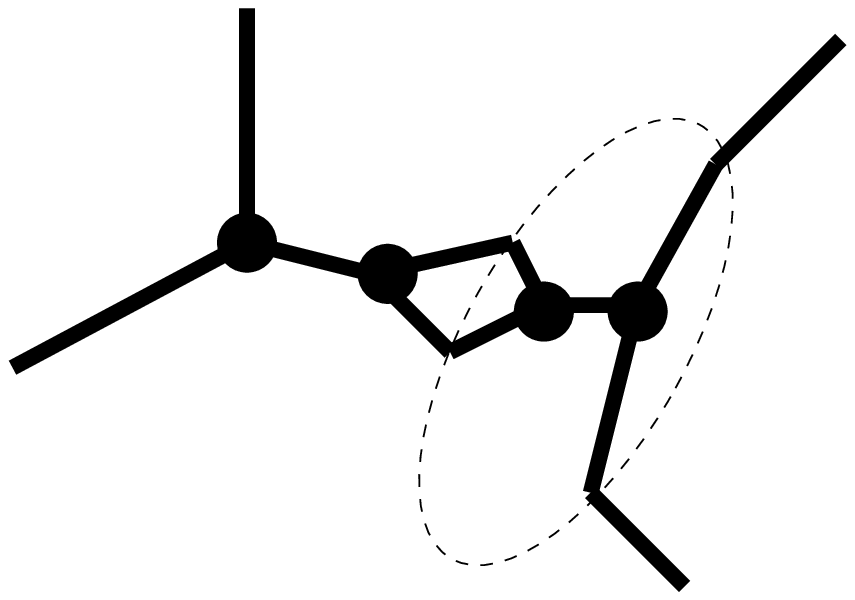}{11}
    \(\to\) \btrackYb{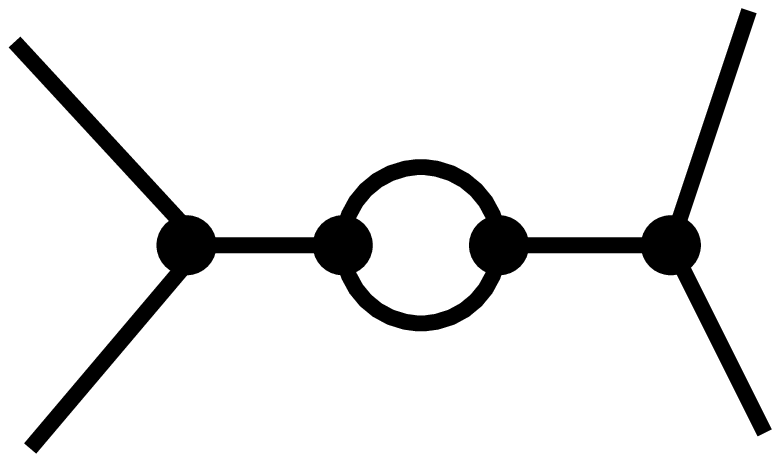}{10}  \(\to\) \btrackYb{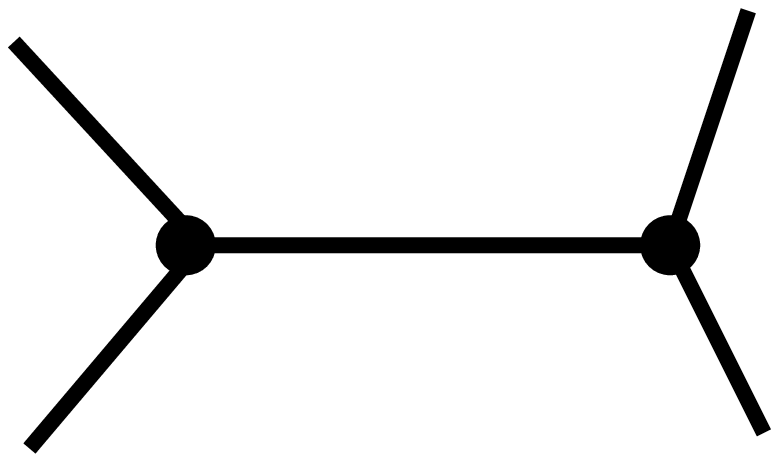}{10}
  \end{center}
  \caption{A reduction of a 4-vertex loop to a sum of ``tree'' tensors,
    weighted by products of 3-\(j\) and 6-\(j\) coefficients.}
  \label{f-recoup}
\end{figure}

Let particles in representations \(\U\) and \(\V\) interact by exchanging a
particle in the representation \(\W\), with the final state particles in the
representations \(\X\) and \(\Z\): \[\btrackYb{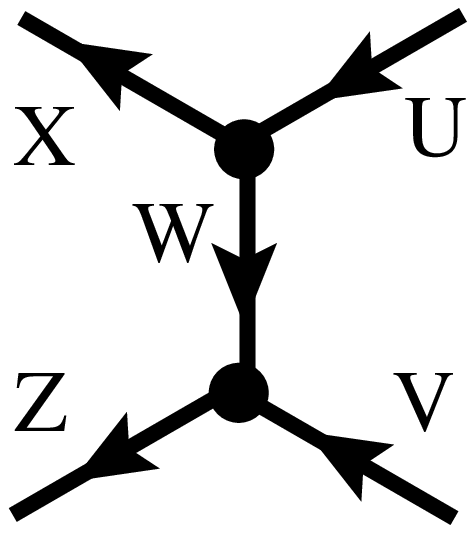}{7}\,.\] Applying the
completeness relation \refeq{Bprojcompl} repeatedly yields
\begin{eqnarray*}
  && \btrackYb{hre.rec1}{7}  
  = \sum_\Y \frac{d_\Y}{\btrackYb{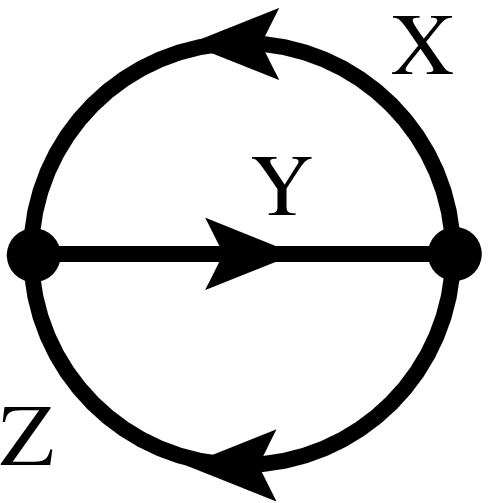}{6}} 
    \; \btrackYb{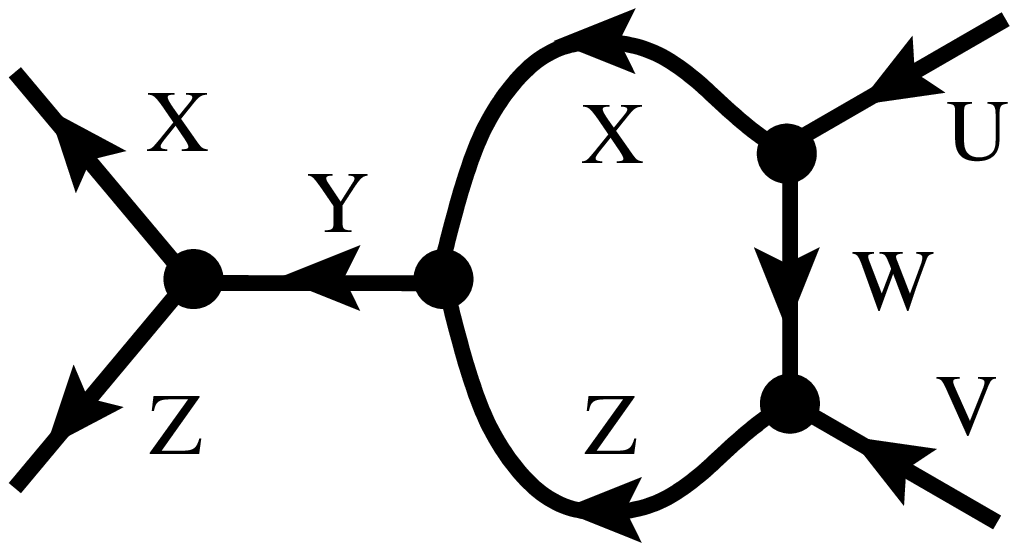}{15} \\
  &&\; = \sum_{\Y, \Y'} 
    \frac{d_\Y d_{\Y'}}{\btrackYb{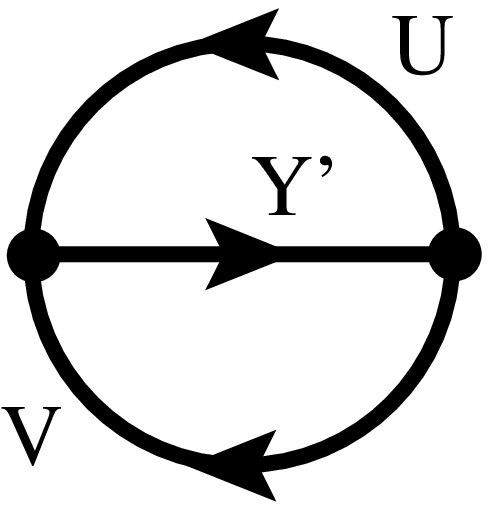}{6}\,
    \btrackYb{hre.rec3}{6}}\, \btrackYb{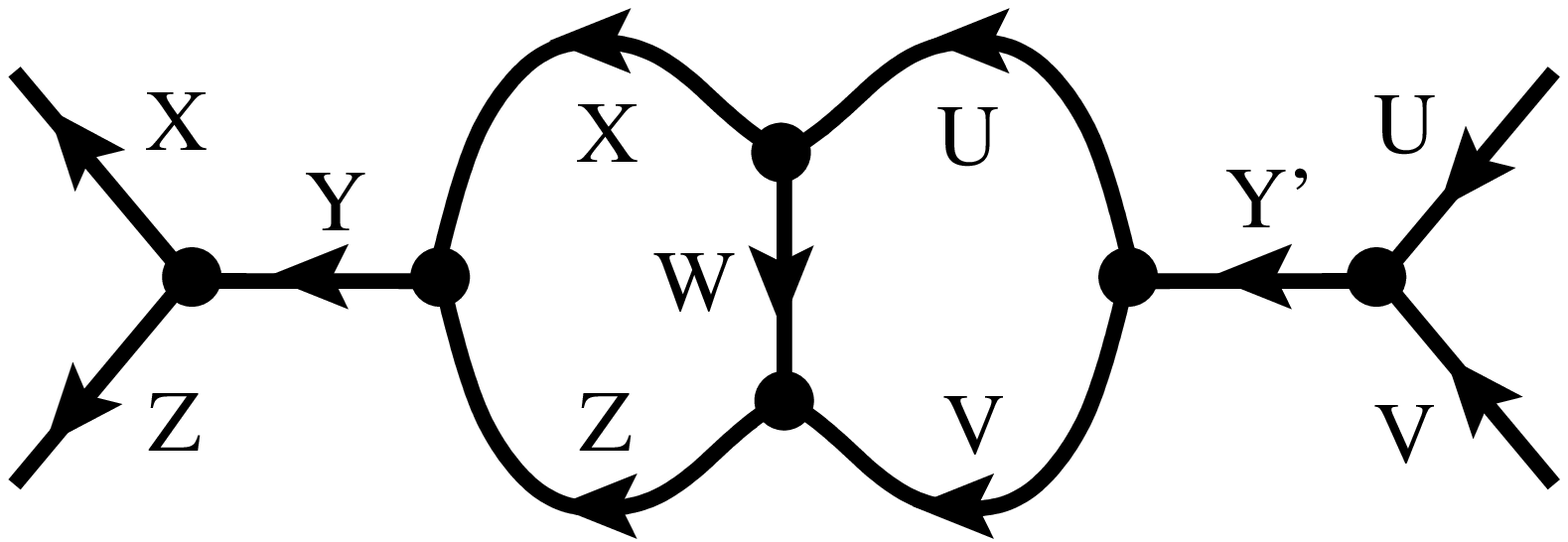}{24}.
\end{eqnarray*}
By the orthogonality of \irrep s \(\Y = \Y'\), and we obtain the
\emph{recoupling relation}
\begin{eqnarray} \label{redrel}
\btrackYb{hre.rec1}{7}
  &=& \sum_{\Y} \; d_\Y \; 
    \frac{\btrackYb{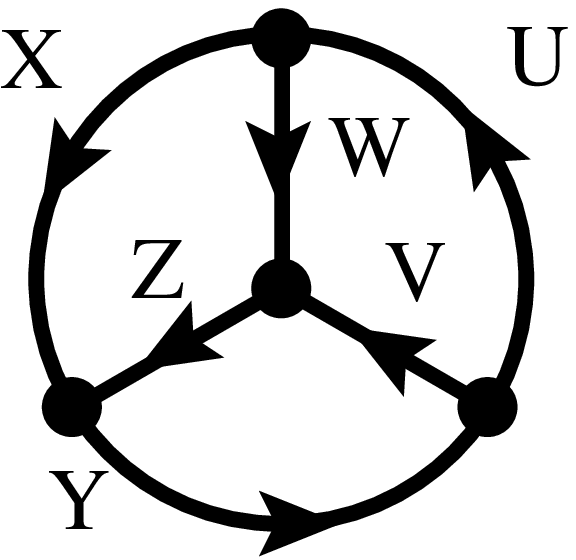}{8}}{\btrackYb{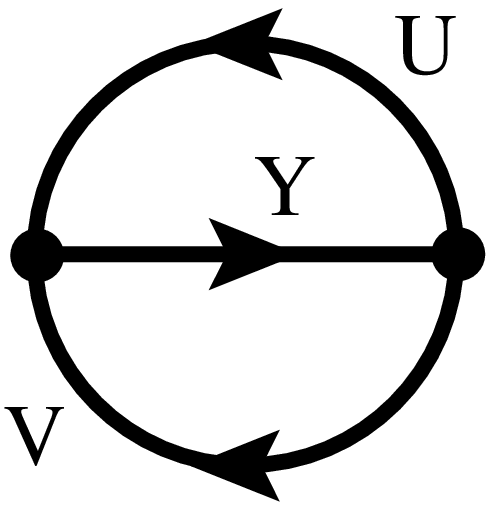}{6}\; 
    \btrackYb{hre.rec3}{6}} \;\; \btrackYb{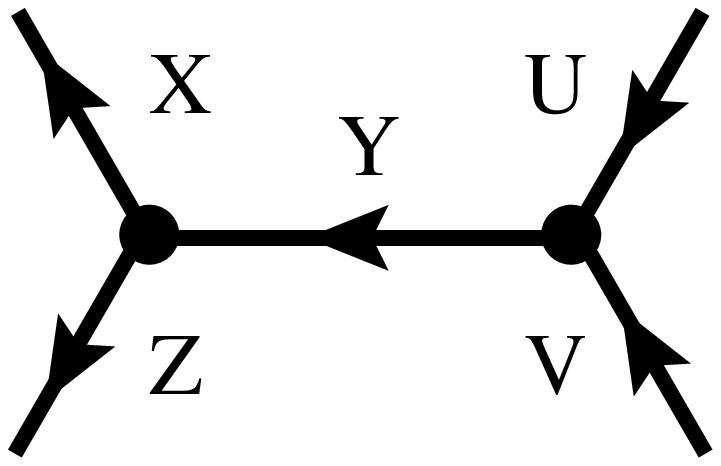}{10} \,.
\end{eqnarray}
The ``Mercedes'' vacuum bubbles in the numerators are called 6-\(j\)
coefficients. 
{\em Any} arbitrarily complicated vacuum bubble can be reduced to 3-\(j\) and
6-\(j\) coefficients by recursive use of the recoupling relation
(\ref{redrel}). For instance, a four vertex loop can be reduced to a two-vertex
loop by repeated application of the recoupling relations as sketched in
\reffig{f-recoup}.

Another, more explicit example of a sequence of recouplings, is the following
step-by-step reduction of a five-particle state:
\begin{eqnarray*}
  \btrackYb{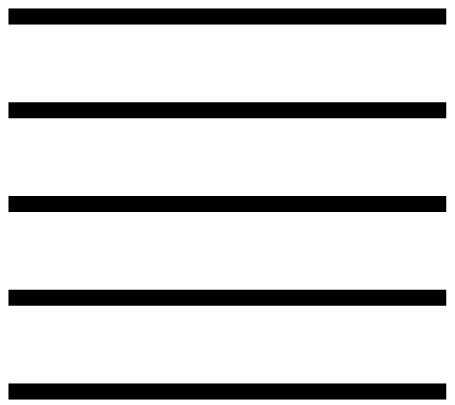}{7} 
  &=& \sum_{\X,\Z} \btrackYb{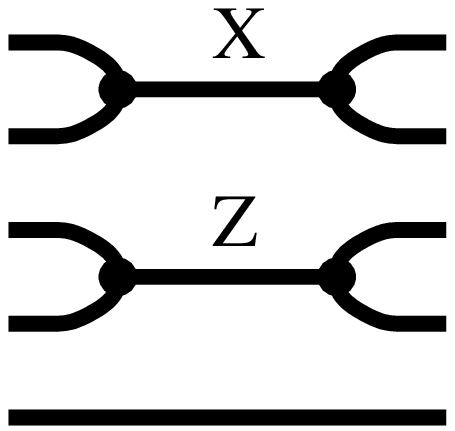}{7} 
  = \sum_{\W,\X,\Z} \btrackYb{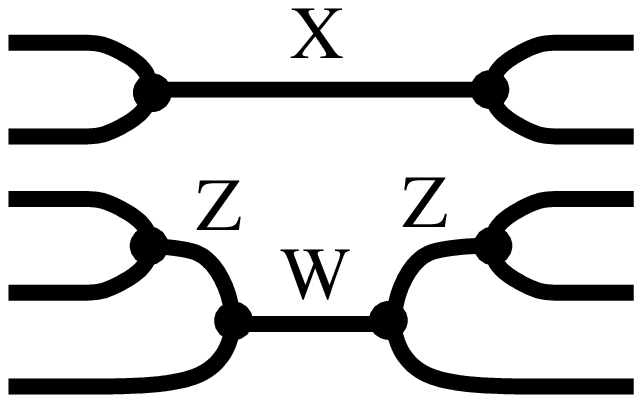}{10} \\[3mm]
  &=& \sum_{\W,\X,\Y,\Z} \btrackYb{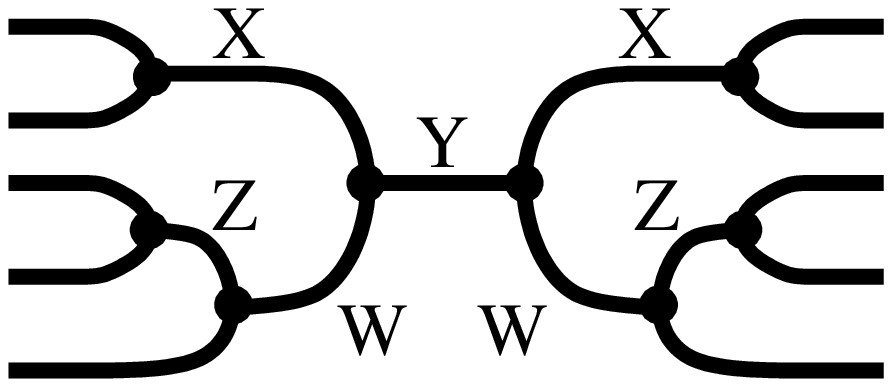}{14} 
\end{eqnarray*}
(for brevity we omit the normalization factors here). Taking the trace of both
sides leads to 12-\(j\) coefficients of the form
\begin{equation}
  \btrackYb{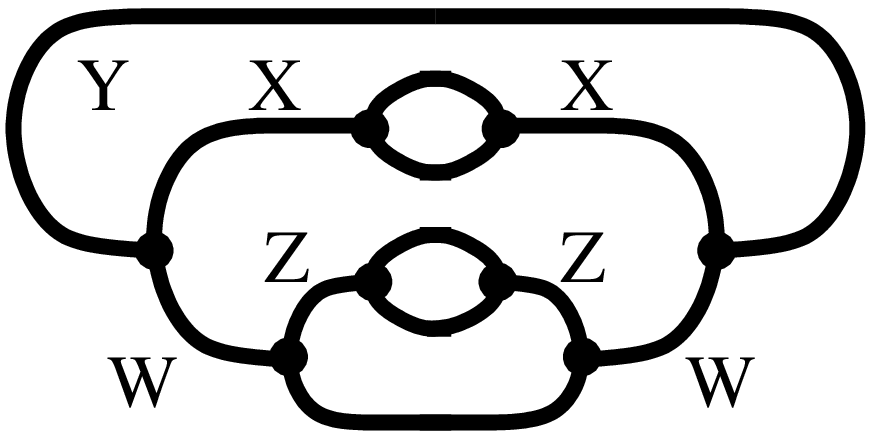}{14}\,.
  \label{examp12j}
\end{equation}

\subsection{\un\ recoupling relations} \label{s:UNrecoupling} 

Due to the overall particle number conservation (we consider no
``anti-particle'' states here), for \un\ the above five-particle recoupling
flow takes a very specific form in terms of \YoungPOps:
\begin{eqnarray*}
  \btrackYb{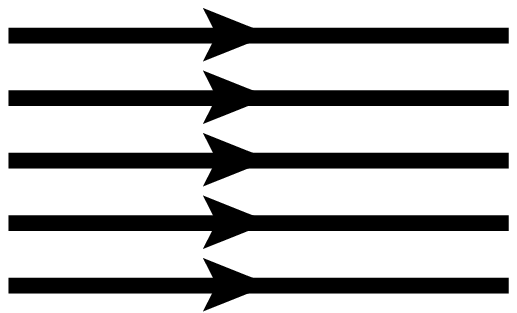}{8} 
  &=& \sum_{\rom{X,Z}} \btrackYb{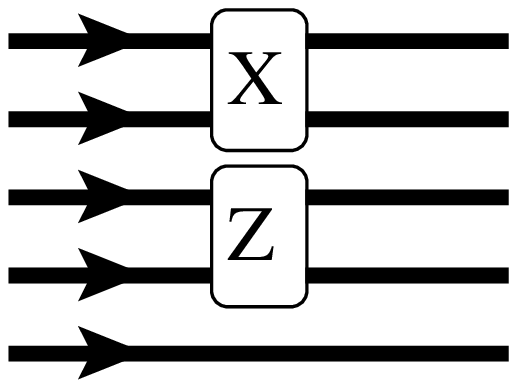}{8} 
  =   \sum_{\rom{W,X,Z}} \btrackYb{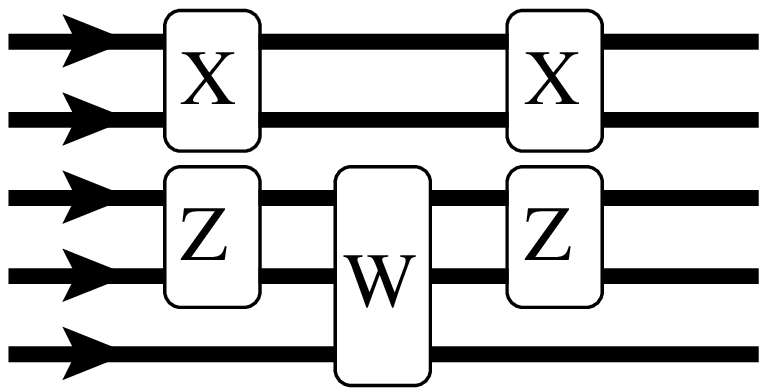}{12} \\[3mm]
  &=&  \sum_{\rom{W,X,\Y,Z}} \btrackYb{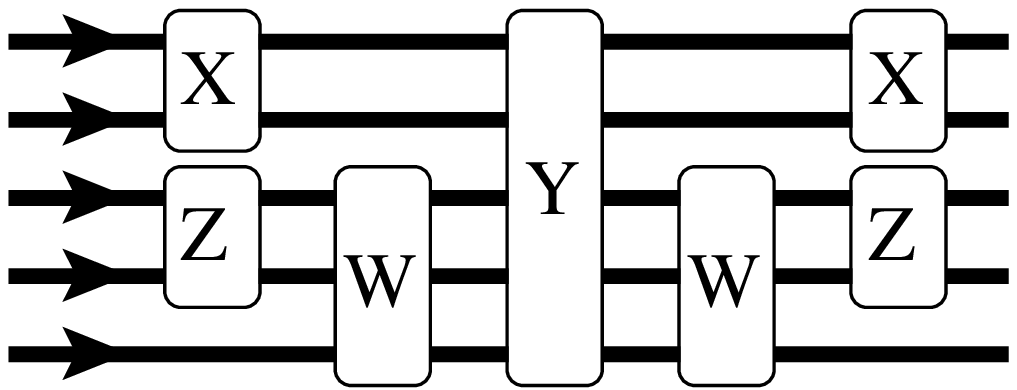}{16} 
\end{eqnarray*}
More generally, we can visualize any sequence of \un\ pairwise Clebsch-Gordan
reductions as a flow with lines joining into thicker and thicker projection
operators, always ending in a maximal \(P_\Y\) which spans across all lines.

In the trace \refeq{examp12j} we can use the idempotency of the projection
operators to double the maximal \YoungPOp\ \(P_\Y\), and sandwich by it
all smaller projection operators:
\begin{eqnarray}
  \btrackYb{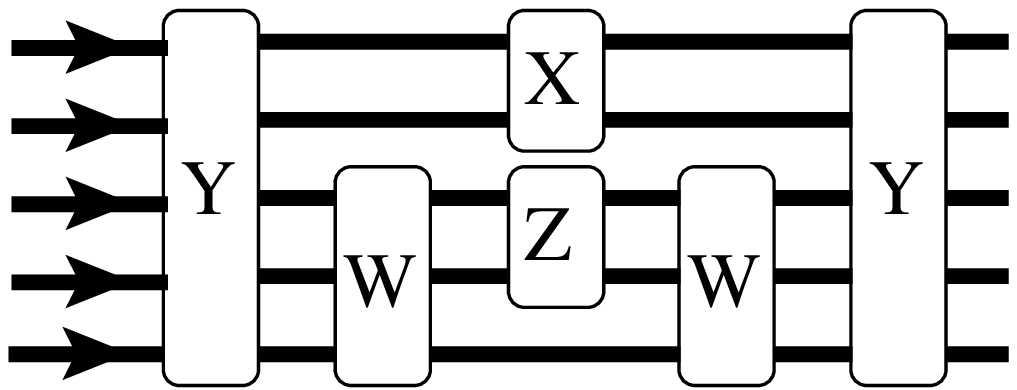}{16}  \,.
  \label{sandwich}
\end{eqnarray}
From uniqueness of the connection between the symmetry operators (see
\refappe{a_unique}) we have for any permutation \(\sigma \in\symgrp_\numBoxes\)
\begin{equation} 
  \btrackYb{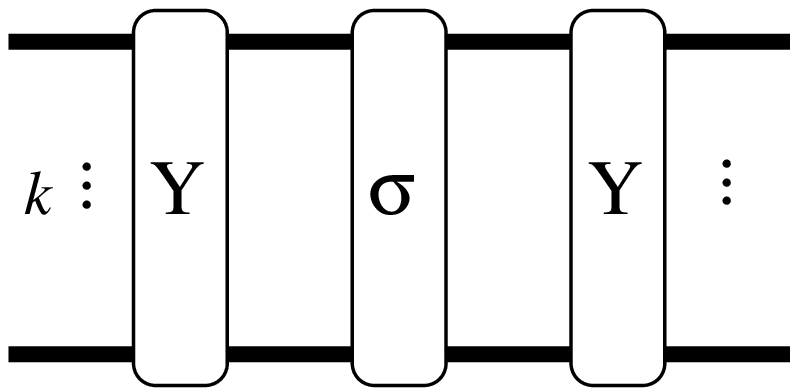}{12} = m_\sigma \, \btrackYb{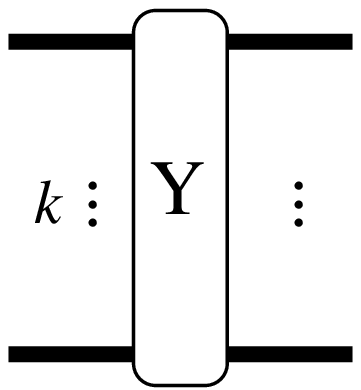}{5}
  \label{msigma}
\end{equation}
where \(m_\sigma = 0, \pm 1\). Expressions like \refeq{sandwich} can be
evaluated by expanding the projection operators \(P_\W\), \(P_\X\),
\(P_\Z\) and determining the value of \(m_\sigma\) of \refeq{msigma} for
each permutation \(\sigma\) of the expansion. The result is
\[\btrackYb{hrecex.1e}{16} = \theconst(\Y;\W,\X,\Z) \btrackYb{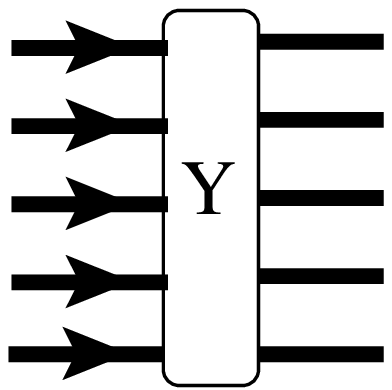}{6}\]
where the factor \(\theconst(\Y;\W,\X,\Z)\) {\em does not depend} on \(n\) and
is determined by a purely symmetric group calculation. Several examples follow.

\subsection{Evaluation of \threenj\ coefficients} \label{s-3j}

Let \(\X\), \(\Y\), and \(\Z\) be \irrep s of \un. In terms of the \YoungPOps\
\(P_\X\), \(P_\Y\), and \(P_\Z\), a \un\ three-vertex \refeq{3vertex} is
obtained by tying together the three \YoungPOps,
\begin{equation}
  \btrackYb{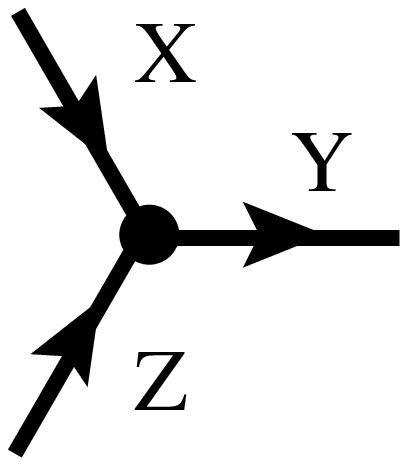}{6} = \btrackYb{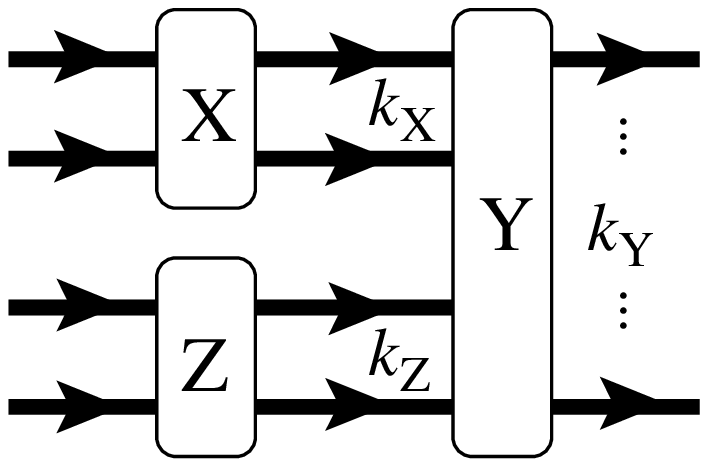}{11} \,.
  \label{defn-3vert}
\end{equation}
The number of particles is conserved (the multi-particle states constructed
here consist only of particles, no ``antiparticles''): \(k_\X + k_\Z = k_\Y\).
A 3-\(j\) coefficient constructed from the vertex \refeq{defn-3vert} is then
\begin{eqnarray} \label{the3j}
  \btrackYb{h3j.1a}{8} \;\; = \;\;\btrackYb{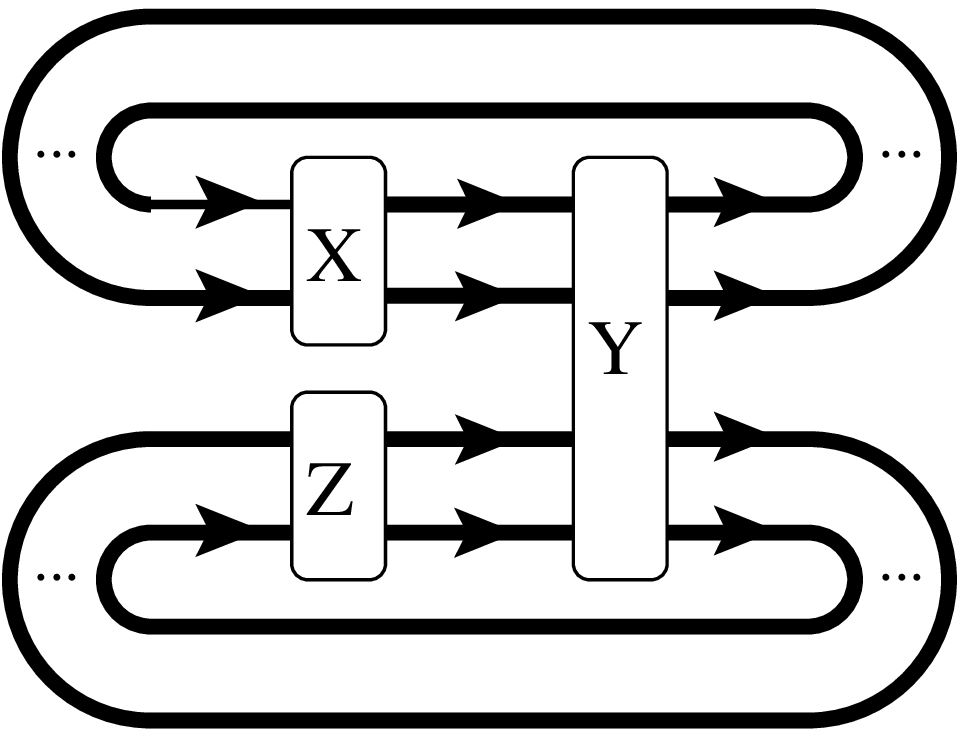}{15} \,.
\end{eqnarray}
As an example, take \[\X = \btrackYb{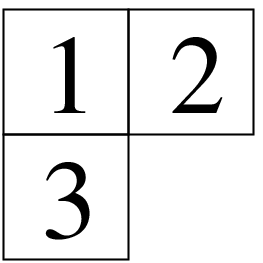}{4} \;, \quad \Y =
\btrackYb{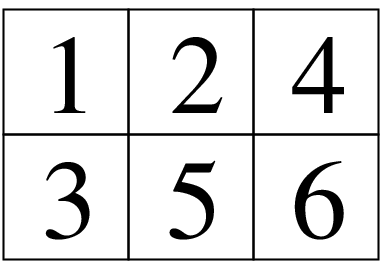}{6} \;, \qquad \rom{and} \qquad \Z =
\btrackYb{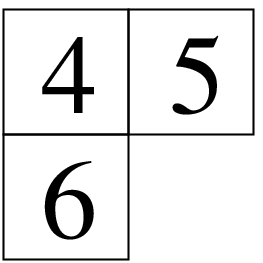}{4}.\] Then
\begin{equation}
  \btrackYb{h3j.1a}{8} \; 
    = \frac{4}{3}\cdot 2 \cdot \frac{4}{3} \btrackYb{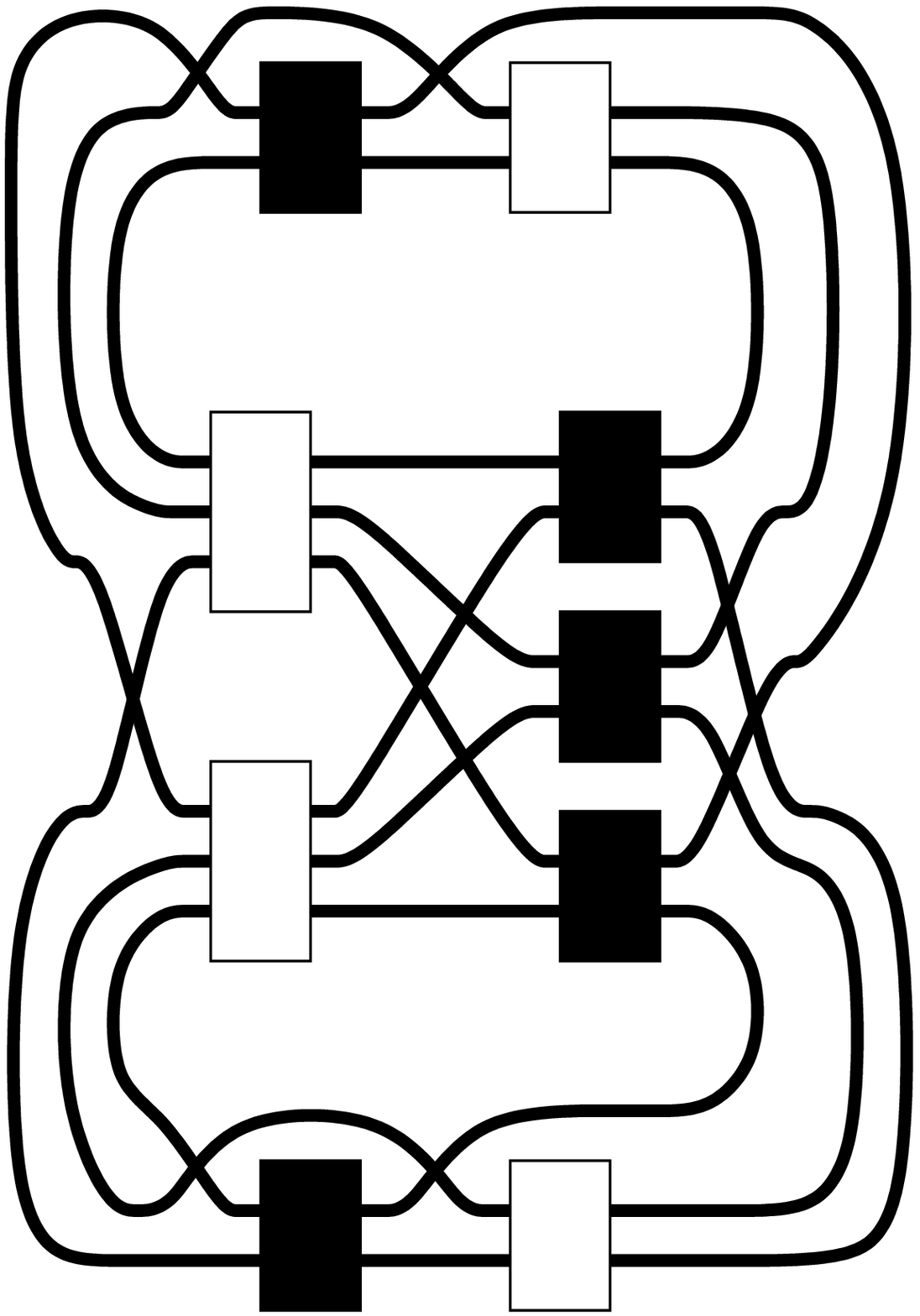}{14} = d_\Y
  \label{3jexample}
\end{equation}
In principle the value of such 3-\(j\) coefficient can be computed by expanding
out all symmetry operators, but that is not recommended as the number of terms
in such expansions grows combinatorially with the total number of boxes in the
Young diagram \(\Y\). Instead, the answer --- in this case \(d_\Y = (n^2-1)n^2
(n+1)(n+2)/144\) --- is obtained as follows.

In general, the 3-\(j\) coefficients \refeq{the3j} can be evaluated by
expanding the projections \(P_\X\) and \(P_\Z\) and determining the value of
\(m_\sigma\) in \refeq{msigma} for each permutation \(\sigma\) of the
expansion.

As an example, consider the 3-\(j\) coefficient \refeq{3jexample}. With
\(P_\Y\) as in \refeq{3jexample} we find
\begin{displaymath}
  \begin{array}{rclccccccc}
    \btrackYb{h3it.nP2}{8} &=&
     \frac14 \Big\{ 
       &\btrackYb{h3it.nid}{4} 
       &-& \btrackYb{hperm.132}{4}
       &+& \btrackYb{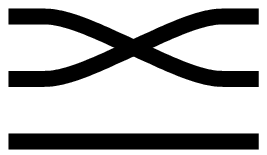}{4} 
       &-& \btrackYb{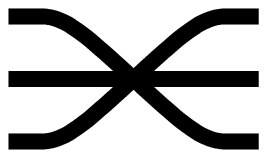}{4}  
     \Big\} \\[3mm]
     m_\sigma(P_\X): & & & +1 & &  0 & & +1 & & -1 ~\\[2mm]
     m_\sigma(P_\Z): & & & +1 & & -1 & & 0  & & -1 \,,
  \end{array} 
\end{displaymath}
hence
\begin{displaymath}
  \btrackYb{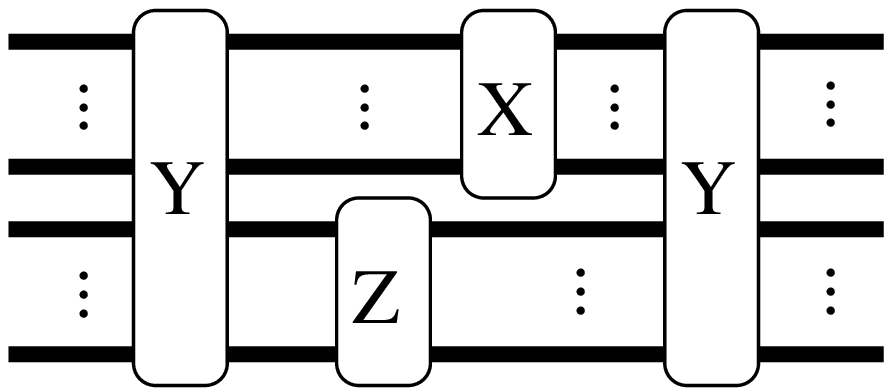}{12} = \left( \frac{3}{4} \right)^2 
    \alpha_\X\alpha_\Z \; \btrackYb{hsandw3}{5}
  = \btrackYb{hsandw3}{5}
\end{displaymath}
and the value of the 3-\(j\) is \(d_\Y\) as claimed in \refeq{3jexample}. That
the eigenvalue happens to be 1 is an accident --- in tabulations of 3-\(j\)
coefficients\rf{Elvang99} it takes a range of values.

The relation \refeq{msigma} implies that the value of any \un\ 3-\(j\)
coefficient \refeq{the3j} is \(\theconst(\Y;\X,\Z) d_\Y\), where \(d_\Y\) is
the dimension of the maximal \irrep~\(\Y\).

A 6-\(j\) coefficient is composed of the three-vertex \refeq{3vertex} and the
other three-vertex in the projection operator \refeq{defn-3vert}, with all
arrows reversed. A general \un\ 6-\(j\) coefficient has form
\begin{equation} 
  \btrackYb{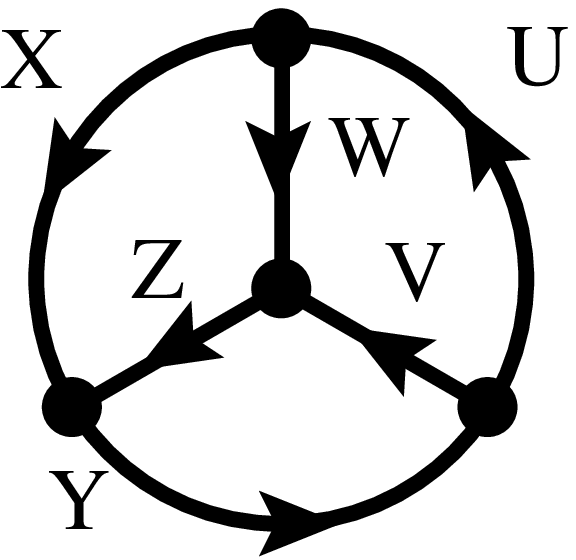}{8} = \btrackYb{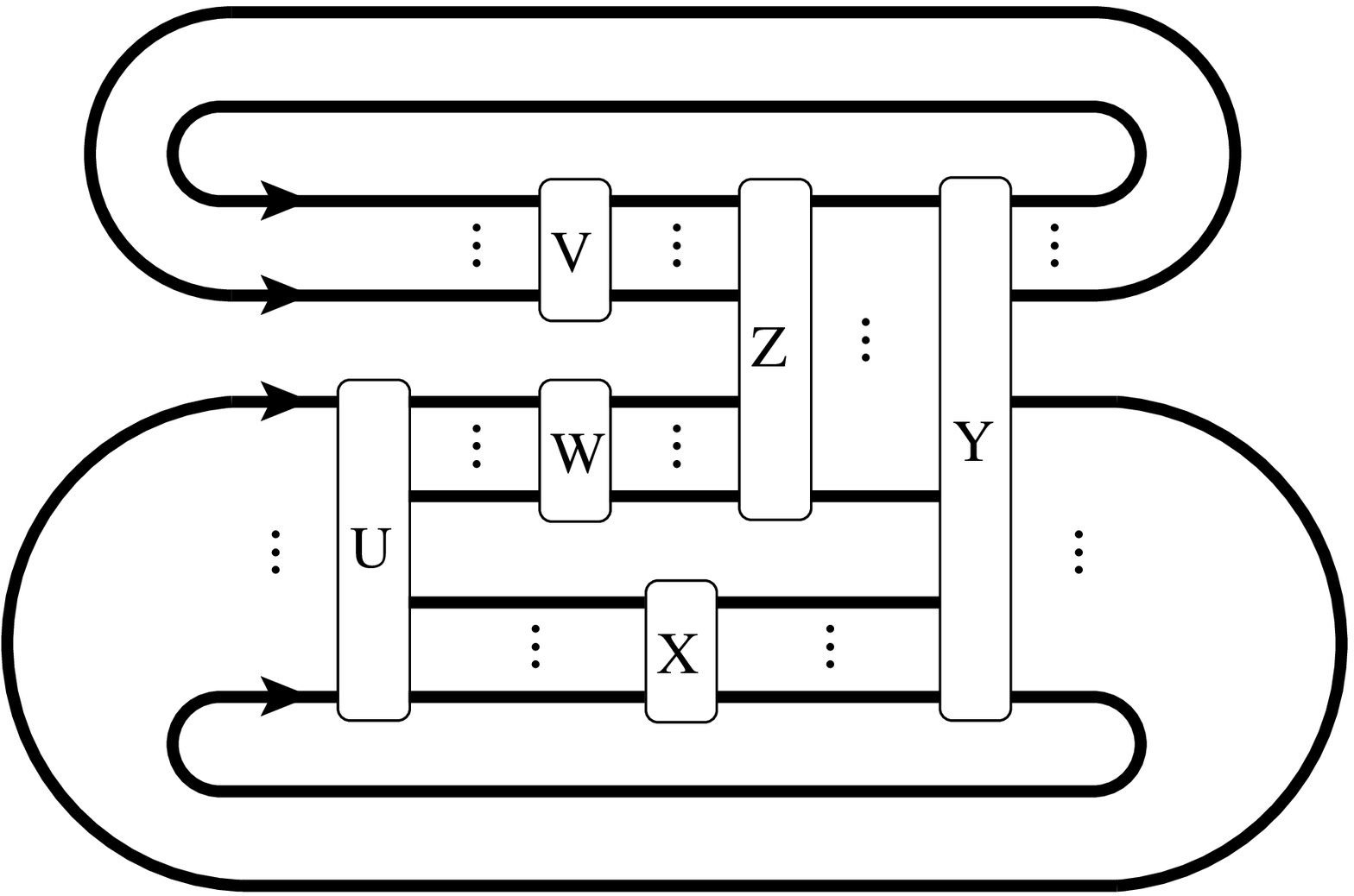}{28}
  \label{the6j}
\end{equation}
Using the relation \refeq{msigma} we immediately see that
\begin{equation}
  \btrackYb{hre.6j.stor}{8} = \theconst \, d_\Y
  \label{value6j}
\end{equation}
where \(\theconst\) is a pure symmetric group \(\symgrp_{k_\Y}\) number,
independent of \un; it is surprising that the only vestige of \un\ is the fact
that the value of a 6-\(j\) coefficient is proportional to the dimension
\(d_\Y\) of its largest projection operator.

\noindent {\bf Example:} Consider the 6-\(j\) constructed from the Young
tableaux
\begin{displaymath}
  \begin{array}{rclcrclcrcl}
    \U &=& \btrackYb{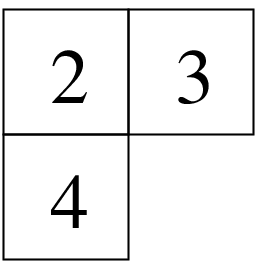}{4} \, ,&&
    \V &=& \btrackYb{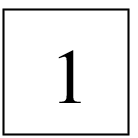}{2} \, ,&&
    \W &=& \btrackYb{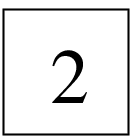}{2} \, , \\
    \X &=& \btrackYb{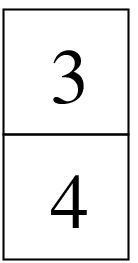}{2} \, ,&&
    \Y &=& \btrackYb{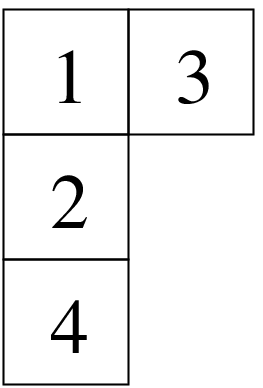}{4} \, ,&&
    \Z &=& \btrackYb{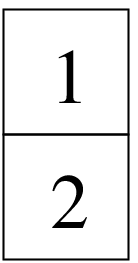}{2} \, .
  \end{array}
\end{displaymath}
Using the idempotency we can double the projection \(P_\Y\) and sandwich the
other operators, as in \refeq{sandwich}. Several terms cancel in the expansion
of the sandwiched operator, and we left with 
\begin{displaymath}
  \begin{array}{rclccccccccc}
  \btrackYb{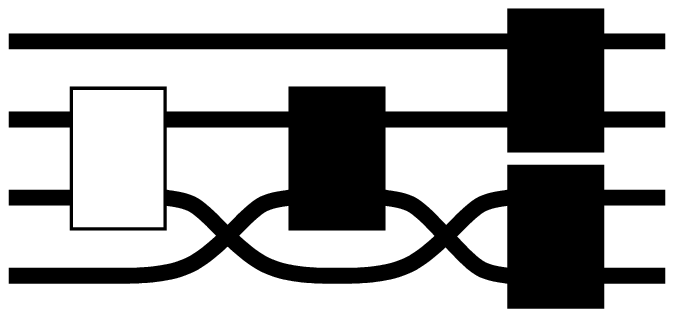}{10}
    &=& \displaystyle \frac1{2^4} \Bigg\{ \btrackYb{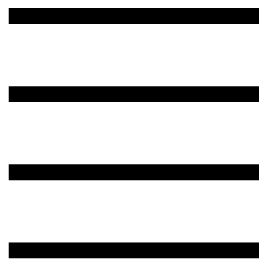}{4}
    &+& \btrackYb{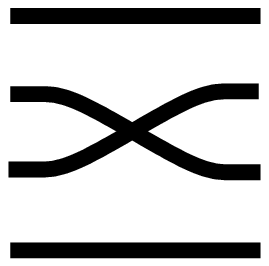}{4} 
    &-& \btrackYb{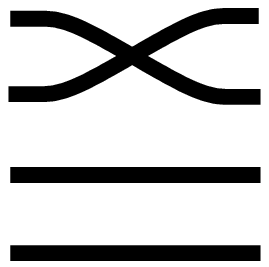}{4}
    &-& \btrackYb{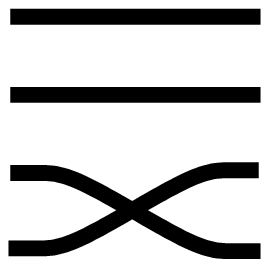}{4} \\[1mm]
    m_\sigma :&& \hspace{8mm}+1 && 0 && -1 && 0 \\[2mm]
    && \hspace{3.3mm}+~\btrackYb{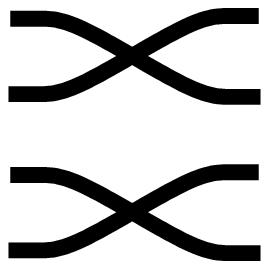}{4}
    &-& \btrackYb{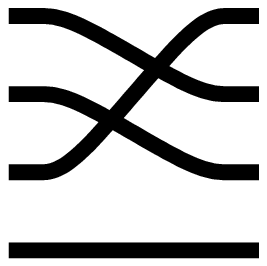}{4}
    &-& \btrackYb{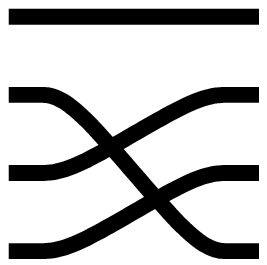}{4}
    &+& \btrackYb{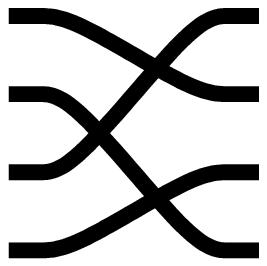}{4} \Bigg\} \\[1mm]
    &&  \hspace{10mm}0 && -1 && 0 && \hspace{-2mm}+1 
    \end{array}
\end{displaymath}
We have listed the symmetry factors \(m_\sigma\) of \refeq{msigma} for each of
the permutations \(\sigma\) sandwiched between the projection operators
\(P_\Y\). We find that in this example the symmetric group factor \(\theconst\)
of \refeq{value6j} is
\begin{eqnarray*}
 \theconst  =  \frac4{2^4}\,
  \alpha_\U\, \alpha_\V\, \alpha_\W\, 
  \alpha_\X\, \alpha_\Z = \frac13 \,,
\end{eqnarray*}
so the value of the 6-\(j\) is
\begin{eqnarray*}
  \btrackYb{hre.6j.stor}{8} 
    = \frac13 d_\Y = \frac{n\,(n^2-1)\,(n-2)}{4!} \,.
\end{eqnarray*}
The method generalizes to evaluations of any {\threenj} coefficients of \un .

\subsection{Sum rules} \label{s:sumrules}

Let \(\Y\) be a standard tableau with \(k_\Y\) boxes, and let \(\Lambda\) be
the set of all standard tableaux with one or more boxes (exclude the trivial
\(\numBoxes=0\) representation). Then the 3-\(j\) coefficients obey the sum
rule
\begin{eqnarray} \label{sumrule}
   \sum_{\X,\Z \in \Lambda} \; \btrackYb{h3j.1a}{6} \quad
   = \quad (k_\Y-1)d_\Y. 
\end{eqnarray}
The sum is finite, because the 3-\(j\) is non-vanishing only if the number of
boxes in X and Z add up to \(k_\Y\), and this happens only for a finite number
of tableaux.

To prove \refeq{sumrule}, recall that the \YoungPOps\ constitute a complete
set, \(\sum_{\X \in \Lambda_\numBoxes} P_\X =\id\), where \(\id\) is the \(k
\times k\) unit matrix and \(\Lambda_\numBoxes\) the set of all standard
tableaux of Young diagrams with \(\numBoxes\) boxes. Hence
\begin{eqnarray*} 
  \sum_{\X,\Z \in \Lambda} & & \btrackYb{h3j.1a}{6} \\
  &=& \sum_{k_\X=1}^{k_\Y-1} 
    \sum_{\begin{array}{c} 
        \scriptstyle \X \in \Lambda_{k_\X} \\[-2pt] 
        \scriptstyle \Z \in \Lambda_{k_\Y-k_\X}
    \end{array}} \;\, \btrackYb{h3j.1b}{15} \\
   &=& \sum_{k_\X=1}^{k_\Y-1} \quad \btrackYb{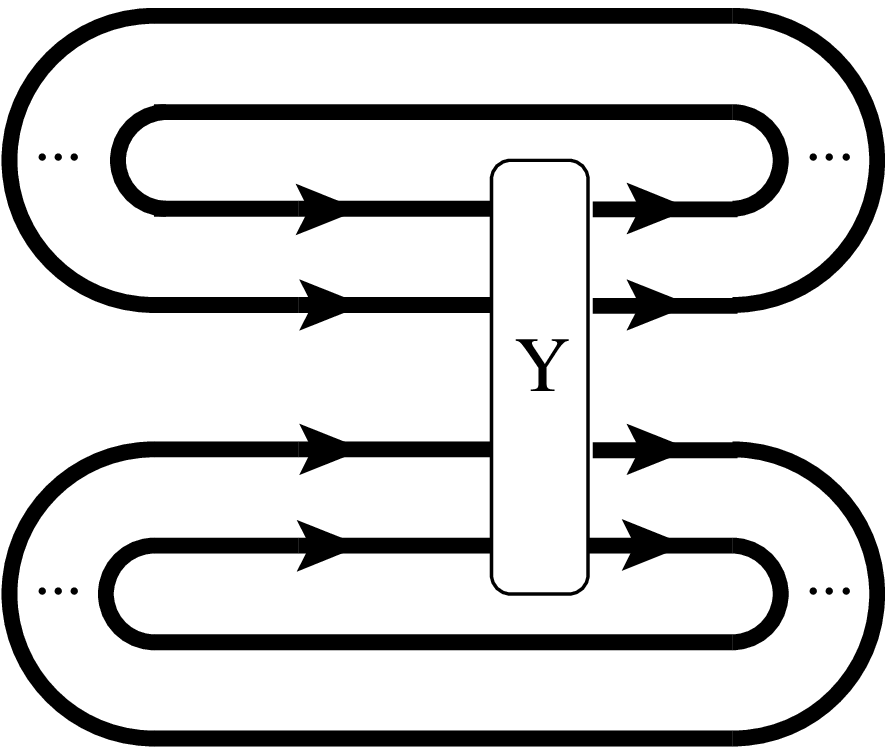}{14} \\
   & = & \; \sum_{k_\X=1}^{k_\Y-1} \; d_\Y
     \;=\; (k_\Y-1)d_\Y \,.
\end{eqnarray*}
This sum rule offers a useful cross-check on tabulations of 3-\(j\) values, see
for instance \refref{Elvang99}.

There is a similar sum rule for the 6-\(j\) coefficients:
\begin{eqnarray} \label{6jsumrule}
  \sum_{\X,\Z,\U,\V,\W \in \Lambda} 
    \btrackYb{hre.6j.stor}{7}
  &=& \frac12(k_\Y-1) (k_\Y-2) \, d_\Y \,.
\end{eqnarray}
Referring to the 6-\(j\) \refeq{the6j}, let \(k_\U\) be the number of boxes in
the Young diagram \(\U\), \(k_\X\) be the number of boxes in \(\X\), etc., and
let \(k_\Y\) be given. From \refeq{the6j} we see that \(k_\rom{X }\) takes
values between \(1\) and \(k_\Y-2\), and \(k_\Z\) takes values between \(2\)
and \(k_\Y-1\), subject to the constraint \(k_\X + k_\Z = k_\Y\). We now sum
over all tableaux U, V, and W keeping \(k_\Y\), \(k_\X\), and \(k_\Z\) fixed.
Note that \(k_\V\) can take values \(1, \dots , k_\Z-1\). Using completeness we
find
\begin{eqnarray*}
  && \hspace{-1.4cm}
    \sum_{\U,\V,\W \in \Lambda} \btrackYb{hre.6j.stor}{7} \\
  &=& \sum_{k_\V=1}^{k_\Z-1} 
    \sum_{\V\in\Lambda_{k_\V}} 
    \sum_{\W\in\Lambda_{k_\Z-k_\V}}
    \sum_{\U\in\Lambda_{k_\Y-k_\V}}
    \btrackYb{hre.6j.stor}{7} \\
  &=& \sum_{k_\V=1}^{k_\Z-1} 
    \btrackYb{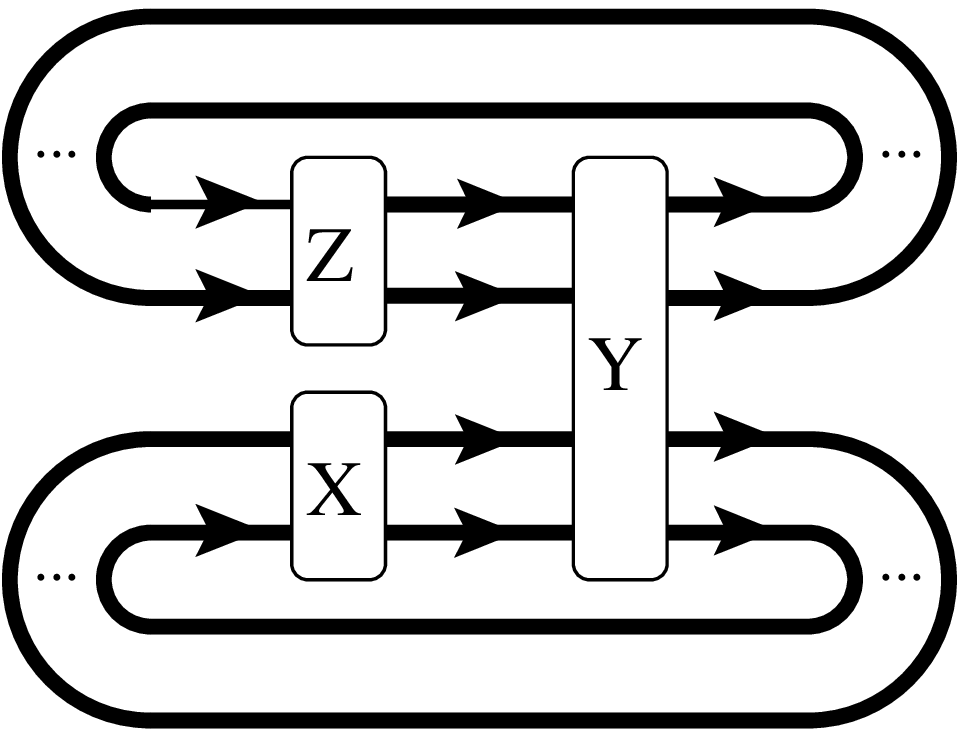}{15} \\
  &=& (k_\Z-1) \, \btrackYb{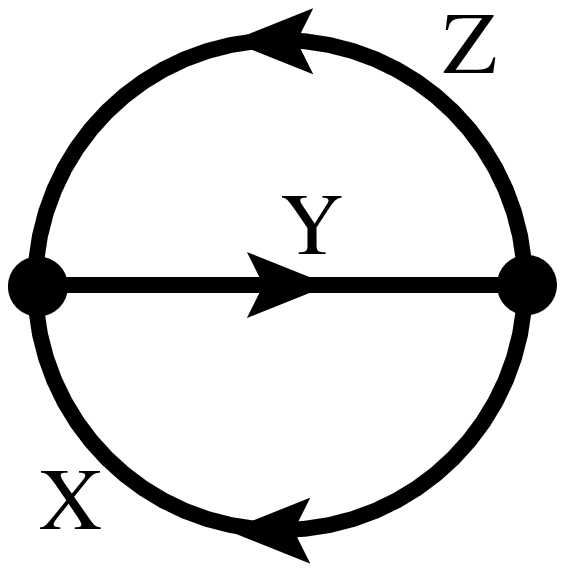}{7} \,.
\end{eqnarray*}
Now sum over all tableaux \(\X\) and \(\Z\) to find
\begin{eqnarray*} 
  && \hspace{-1.6cm} \sum_{\X,\Z,\U,\V,\W \in \Lambda}
    \btrackYb{hre.6j.stor}{7} \\
  &=& \sum_{k_\Z=2}^{k_\Y-1} (k_\Z-1)
    \sum_{\Z\in \Lambda_{k_\Z}}
    \sum_{\X\in \Lambda_{k_\Y - k_\Z}}
    \btrackYb{h3j.in6j.sr}{7} \\
  &=& \frac12(k_\Y-1) (k_\Y-2) \, d_\Y
\end{eqnarray*}
verifying the sum rule \refeq{6jsumrule} for 6-\(j\) coefficients.

\section{\sun\ and its adjoint representation} \label{s-adjoint}

The \sun\ group elements satisfy \(\det U = 1\), so \sun\ has an additional
invariant, the Levi-Civita tensor 
\[
\varepsilon_{a_1a_2\dots a_n} =
U{}_{a_1}{}^{a'_1}U{}_{a_2}{}^{a'_2} \cdots U{}_{a_n}{}^{a'_n} 
\varepsilon_{a'_1a'_2\dots a'_n} 
\,.\] 
In the diagrammatic notation the Levi-Civita tensors can be drawn
as\rf{PenroseMacCullen}
\begin{displaymath}
  \frac1{\sqrt{n!}} \, \varepsilon_{a_1 a_2 \dots a_n} = \btrackYb{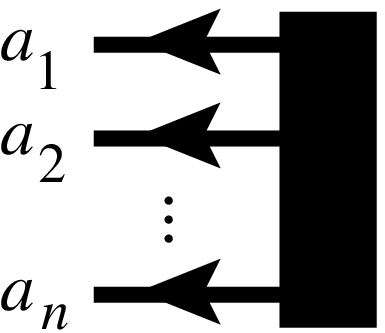}{6}
  \,,\quad 
  \frac1{\sqrt{n!}} \, \varepsilon^{a_n \dots a_2 a_1} = \btrackYb{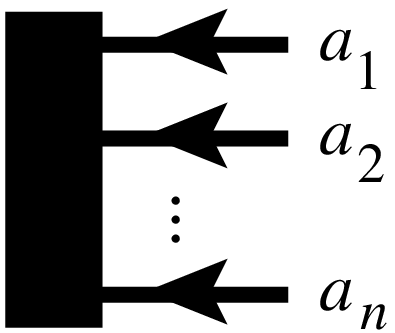}{6}
  \,.
\end{displaymath}
They satisfy
\begin{eqnarray}
  \btrackYb{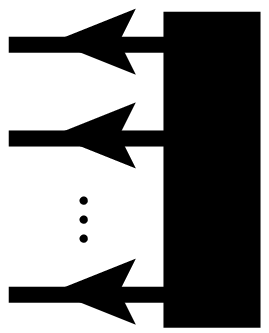}{4} \, \btrackYb{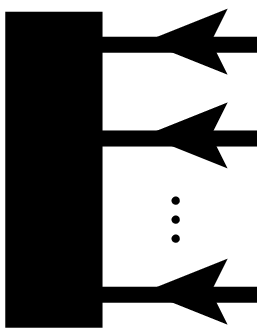}{4} = \btrackYb{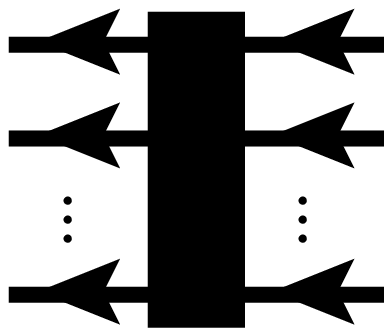}{6} 
   \label{lc-to-A}
\end{eqnarray}
(Levi-Civita projects an \(n\)-particle state onto a single, 1-dimensional,
singlet representation), and are correctly normalized,
\begin{eqnarray*}
  \btrackYb{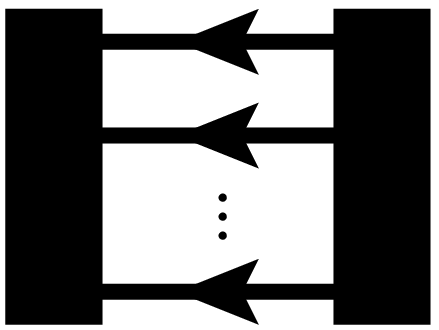}{7} = 1 \,.
\end{eqnarray*}

The Young diagrams for \sun\ cannot contain more than \(n\) rows, since at most
\(n\) indices can be antisymmetrized. By contraction with the Levi-Civita
tensor, a column with \(k\) boxes can be converted into a column of \(n-k\)
boxes: this operation associates to each \irrep\ the \emph{conjugate} \irrep.
The Young diagram corresponding to the \irrep\ is the \emph{conjugate} Young
diagram constructed from the missing pieces needed to complete the rectangle of
\(n\) rows. For example, the conjugate of the \irrep\ corresponding to the
partition \([4,2,2,1]\) of \(SU(6)\) has the partition \([4,4,3,2,2]\):
\[SU(6):~~\btrackYb{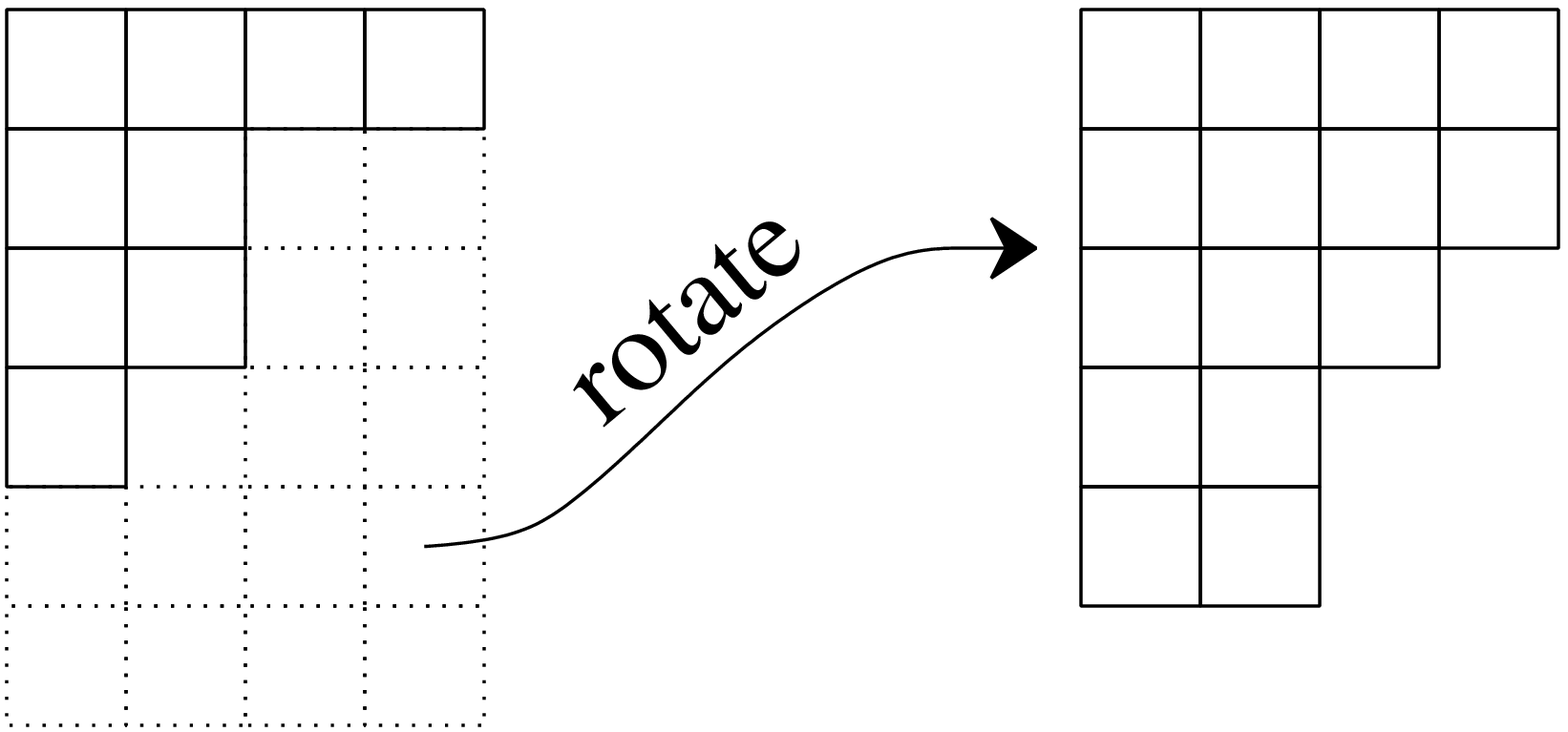}{26} \,.\] 

The Levi-Civita tensor converts an antisymmetrized collection of \(n-1\) ``in''
indices, an \((n-1)\)-particle state, into 1~``out'' index: a single
anti-particle state \(\bar{\btrackYb{Ybox}{2}}\), the conjugate of the
fundamental representation \btrackYb{Ybox}{2} single particle state. The
corresponding Young diagram is a single column of \(n-1\) boxes. The product of
the fundamental representation and the conjugate representation of \sun\
decomposes into a singlet and the adjoint representation:
\begin{displaymath}
  \begin{array}{rclcccl}
    \btrackYb{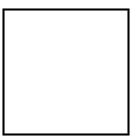}{2} &\otimes& 
      \left.\btrackYb{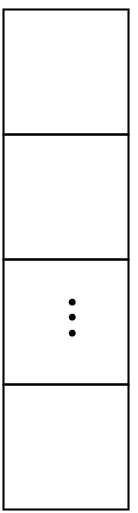}{2}\right\}\scriptstyle{n-1} 
    & = & 1 & ~\oplus &  
    ~\left.\btrackYb{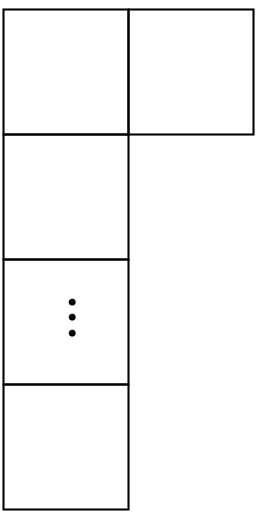}{4}\right\}\scriptstyle{n-1} \\[6mm]
      n\, & \cdot & ~n & = & 1 & + & (n^2 -1) \,.
  \end{array}
\end{displaymath}
In the notation introduced in \refsect{YoungPOps}, the \YoungPOp\ for the
adjoint representation \(A\) is drawn as
\begin{displaymath}
  P_A = \frac{2 \, (n-1)}{n} \; \btrackYb{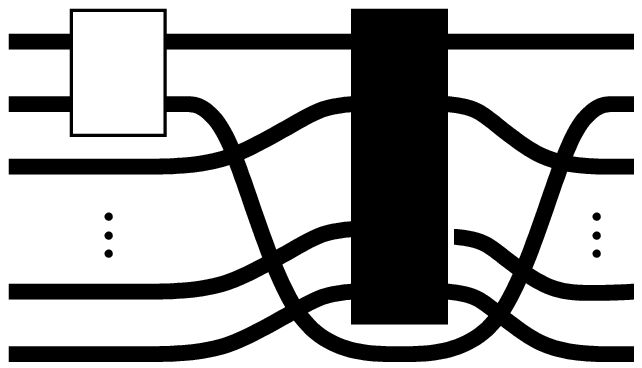}{10}\,. 
\end{displaymath}
Using \(P_A\) and the definition \refeq{defn-3vert} of the three-vertex, \sun\
group theory weights involving quarks, antiquarks, and gluons can be calculated
by expansion of the symmetry operators or by application of the recoupling
relation. 
When the adjoint representation plays a key role, as it
does in gauge theories, it is wisest to abandon the above construction of all
\irrep s by Clebsch-Gordan reductions of multi-particle states, and build the
theory by taking a single particle and a single anti-particle as the
fundamental building blocks. A much richer theory, beyond the scope of this
article, follows, leading to a diagrammatic construction of representations of
all simple Lie groups, the classical as well as the exceptional. The reader is
referred to \refref{PCgr} for the full exposition. 

\section{Negative dimensions} \label{s-negdim}

We conclude by a brief discussion of implications of the \(n \to -n\)
duality\rf{PC-ADK,PCgr} of \un\ invariant scalars.

Any \sun\ invariant tensor is built from Kronecker deltas and Levi-Civita
tensors. A scalar is a tensor object with all indices contracted, so in the
diagrammatic notation a scalar is a diagram with no external legs, a vacuum
bubble. Thus, in scalars Levi-Civita tensors can appear only in pairs (the
lines must end somewhere), and by \refeq{lc-to-A} the Levi-Civita tensors
combine to antisymmetrizers. Consequently both \un\ and \sun\ invariant scalars
are all built only from symmetrizers and antisymmetrizers.

Expanding all symmetry operators in a \un\ vacuum bubble gives a sum of
entangled loops. Each loop is worth \(n\), so each term in the sum is a power
of \(n\), and therefore a \un\ invariant scalar is a polynomial in \(n\).

The negative dimensionality theorem\rf{PC-ADK,PCgr} for \un\ states that
interchanging symmetrizers and antisymmetrizers in a \un\ invariant scalar is
equivalent (up to an overall sign) to substituting \(n \to -n\) in the
polynomial, which is the value of the scalar. We write this \[\overline{{U}(n)}
= {U}(-n).\] The bar symbolizes the interchange of symmetrizers and
antisymmetrizers.

The terms in the expansion of all symmetry operators in a \un\ vacuum bubble
can be arranged in pairs that only differ by one crossing,
\begin{equation} \label{nd1} 
  \dots + \btrackYb{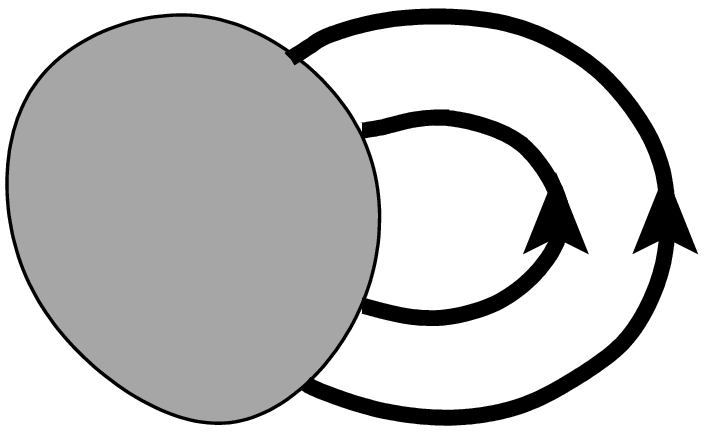}{11} \pm \btrackYb{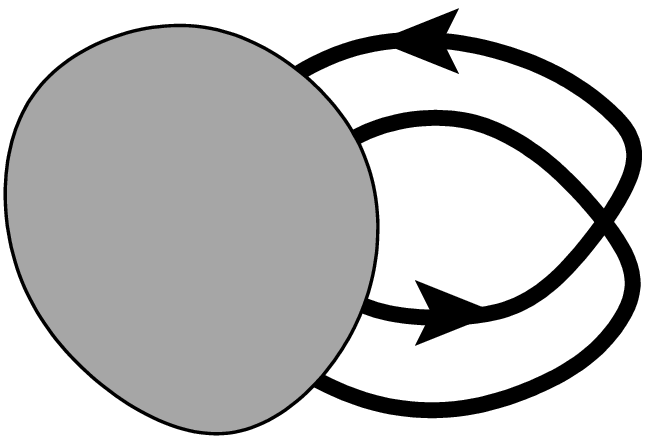}{10} + \dots \,,
\end{equation}
with \(\pm\) depending on whether the crossing is due to symmetrization (\(+\))
or antisymmetrization (\(-\)). The gray blobs symbolize the tangle of lines
common to the two terms.

If the two arcs outside the gray blob of first term of \refeq{nd1} belong to
separate loops, then in the second term they will belong to the same loop. The
two terms thus differ only by a factor of \(n\): schematically,
\[\btrackYb{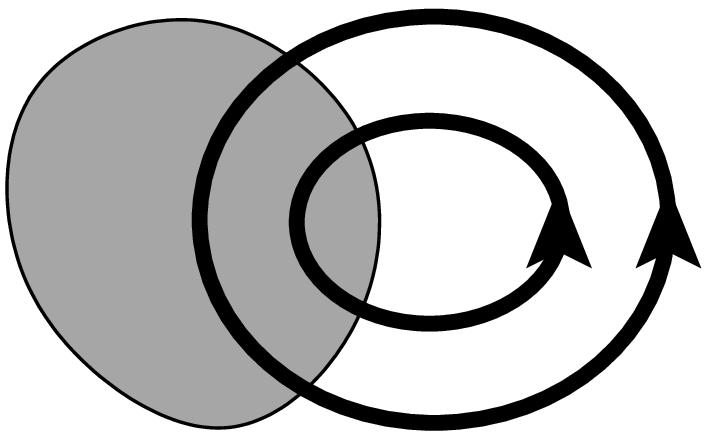}{11} = n\; \btrackYb{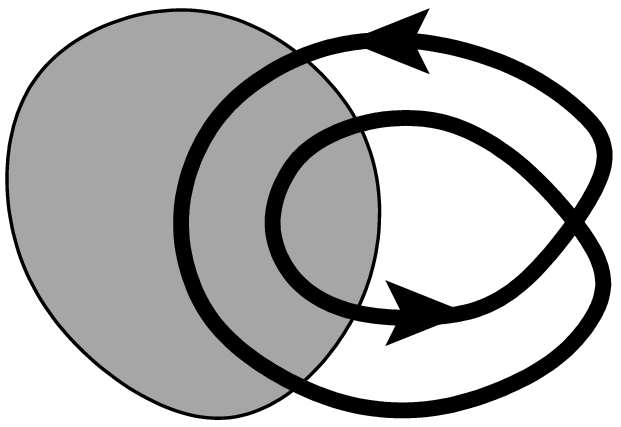}{10} \,.\] Likewise, if
the arcs in the first term belong to the same loop then in the second term they
will belong to two separate loops. In this case the first term is \(1/n\) times
the second term. In either case the ratio of the two terms is an odd power
of~\(n\). Interchanging symmetrizers and antisymmetrizers in a \un\ vacuum
bubble changes the sign in \refeq{nd1}. Up to an overall sign the result is the
same as substituting \(n \to -n\). This proves the theorem.

Consider now the implications for the dimension formulas and the values of
\threenj\ coefficients. The dimension of an \irrep\ of \un\ is the trace of the
\YoungPOp, a vacuum bubble diagram built from symmetrizers and
antisymmetrizers. Applying the negative dimensionality theorem we get
\(d_{\Y^t}(n) = d_\Y(-n)\), where \(\Y^t\) is the \emph{transpose} \(\Y^t\) of
the standard Young tableau \(\Y\) obtained by interchanging rows and columns
(reflection across the diagonal). For instance \([3,1]\) is the transpose of
\([2,1,1]\), \[\left(\btrackYb{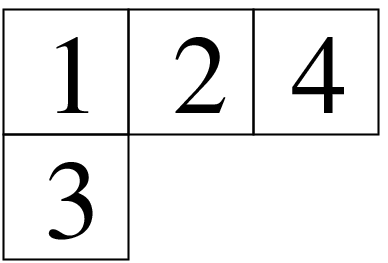}{6}\right)^t = \;\;
\btrackYb{h4it.Y8}{4} \,.\] As an example, note the \(n \to -n\) dualities in
the dimension formulas of \reffig{red3it}.

Now for standard tableaux \(\X\), \(\Y\), and \(\Z\), compare the diagram of
the 3-\(j\) constructed from \(\X\), \(\Y\), and \(\Z\) to that constructed
from \(\X^t\), \(\Z^t\), and \(\Y^t\). The diagrams are related by a reflection
in a vertical line, reversal of all the arrows on the lines, and interchange of
symmetrizers and antisymmetrizers. The first two operations do not change the
value of the diagram, hence the values of the two diagrams are again related by
\(n \leftrightarrow -n\) (and possibly an overall sign; this sign is fixed by
requiring that the highest power of \(n\) comes with a positive coefficient).
Hence in tabulation it is sufficient to calculate approximately half of all
3-\(j\)'s. The 3-\(j\) sum rule \refeq{sumrule} provides a cross-check.

The two 6-\(j\) coefficients \[\btrackYb{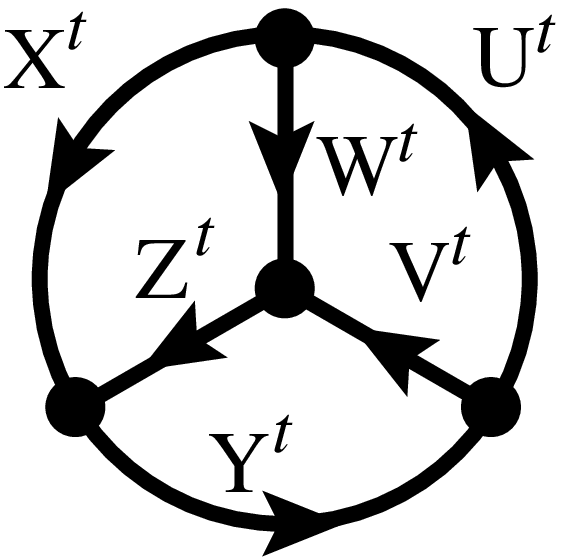}{7}\hspace{4em}
\btrackYb{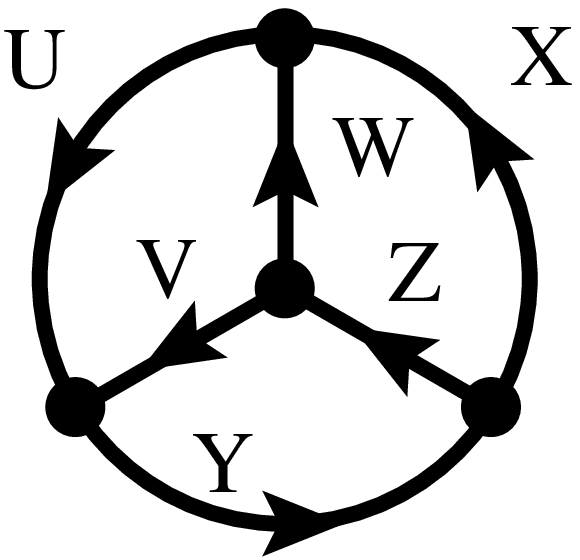}{7}\] are related by a reflection in a vertical line,
reversal of all the arrows on the lines, and interchange of symmetrizers and
antisymmetrizers --- this can be seen by writing out the 6-\(j\) coefficients
in terms of the \YoungPOps\ as in \refeq{the6j}. By the negative dimensionality
theorem, the values of the two 6-\(j\) coefficients are therefore again related
by \(n \leftrightarrow -n\).

\section{Summary} \label{s:concl}

We have presented a diagrammatic method for construction of correctly
normalized \YoungPOps\ for \un. These projection operators in diagrammatic form
are useful for explicit evaluation of group theoretic quantities such as the
{\threenj} coefficients. Using the recoupling relations, all \un\ invariant
scalars can be reduced to expressions involving only terms of 3-\(j\) and
6-\(j\) coefficients and the dimensionalities of the representations. Our main
results are:
\begin{itemize}
\item Diagrammatic \YoungPOps\ for tensors (multi-particle states) with given
  symmetry properties; a diagrammatic proof of their uniqueness, completeness
  and orthogonality.
\item \un\ invariant scalars may be expressed in terms of the \YoungPOps, and
  their values computed by diagrammatic expansions.
\item A strand-diagram proof of the dimension formula for the \irrep s of \un.
\item \un\ 3-\(j\) and 6-\(j\) coefficients constructed from the three-vertex
  defined in \refeq{defn-3vert} have simple \(n\)-dependencies: they are
  proportional to the dimension of the maximal \irrep\ projection operator that
  spans over all multi-particle indices.
\item The negative dimensionality theorem applies to all \un\ invariant
  scalars, in particular the {\threenj} coefficients and the dimensions of the
  \irrep s of \un.
\item The sum rules \refeq{sumrule} and \refeq{6jsumrule} for 3-\(j\) and
  6-\(j\) coefficients afford useful cross-checks of {\threenj} tabulations.
\end{itemize}

\subsection*{Acknowledgements}
HE would like to thank the
University of Edinburgh, UK, the Niels Bohr Institute, Copenhagen, and the
Center for Nonlinear Studies, Georgia Tech, for hospitality, and the Danish
Research Council for support. We thank M.~Hermele for bringing \refref{Sagan01}
to our attention.
 
\appendix

\section{Diagrammatic \YoungPOps: the proofs} \label{appendixA}

In this appendix we prove the properties of the \YoungPOps\ stated above in
\refsect{YoungPOps}.

\subsection{Uniqueness} \label{a_unique}

We show that the \YoungPOps\ \(P_\Y\) are well-defined by proving the existence
and uniqueness (up to an overall sign) of a non-vanishing connection between
the symmetrizers and antisymmetrizers in \(P_\Y\).

\begin{figure}[thb]
  \[\btrackYb{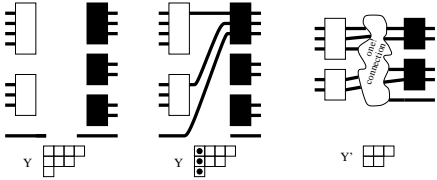}{33}\]
  \caption{There is a unique (up to an overall sign) connection between the
    symmetrizers and the antisymmetrizers, so the \YoungPOps\ are well-defined
    by the construction procedure explained in the text. The figure shows the
    principles of the proof. The dots on the middle Young diagram mark boxes
    that correspond to contracted lines.}
  \label{f-unique}
\end{figure}

The proof is by induction over the number of columns \(t\) in the Young diagram
\(\Y\); the principles are illustrated in \reffig{f-unique}. For \(t=1\) the
\YoungPOp\ consists of one antisymmetrizer of length \(s\) and \(s\)
symmetrizers of length 1, and clearly the connection can only be made in one
way, up to an overall sign, see \reffig{SA-prop}(b).

Assume the result to hold for \YoungPOps\ derived from Young diagrams with
\(t-1\) columns. Let \(\Y\) be a Young diagram with \(t\) columns. The lines
from A\(_1\) in \(P_\Y\) must connect to different symmetrizers for the
connection to be non-zero. Since there are exactly \(|\col_1|\) symmetrizers in
\(P_\Y\), this can be done in essentially one way, since which line goes to
which symmetrizer is only a matter of an overall sign, and where a line enters
a symmetrizer is irrelevant due to \reffig{SA-prop}(a).

After having connected A\(_1\), connecting the symmetry operators in the rest
of \(P_\Y\) is the problem of connecting symmetrizers to antisymmetrizers in
the \YoungPOp\ \(P_{\Y'}\), where \(\Y'\) is the Young diagram obtained from
\(\Y\) by slicing off the first column. Thus \(\Y'\) has \(k-1\) columns, so by
the induction hypothesis the rest of the symmetry operators in \(P_\Y\) can be
connected in exactly one non-vanishing way (up to an overall sign).

\subsection{Orthogonality} \label{a_ortho}

If \(\Y_a\) and \(\Y_b\) denote standard tableaux derived from the same Young
diagram \(\Y\), then \(P_{\Y_a}P_{\Y_b} = P_{\Y_b} P_{\Y_a} = \delta_{ab}
P_{\Y_a}^2\), since there is a permutation of the lines connecting the symmetry
operators of \(\Y\) with those of Z and by uniqueness of the non-zero
connection the result is either \(P_{\Y_a}^2\) (if \(\Y_a=\Y_b\)) or \(0\) (if
\(\Y_a\ne\Y_b\)).

Next, consider two different Young diagrams \(\Y\) and Z with the same number
of boxes. Since at least one column must be bigger in (say) \(\Y\) than in Z
and the \(p\) lines from the corresponding antisymmetrizer must connect to
different symmetrizers, it is not possible to make a non-zero connection
between the antisymmetrizers of \(P_{\Y_a}\) to the symmetrizers in
\(P_{\Z_b}\), where subscripts \(a\) and \(b\) denote any standard tableaux of
\(\Y\) and Z. Hence \(P_{\Y_a}P_{\Z_b} = 0\), and by a similar argument,
\(P_{\Z_b}P_{\Y_a} = 0\).

\subsection{Normalization and completeness} \label{a_norm}

We now derive the formula for the normalization factor \(\alpha_\Y\) such that
the \YoungPOps\ are idempotent, \(P_{\Y_a}^2 = P_{\Y_a}\). By the normalization
of the symmetry operators, \YoungPOps\ corresponding to fully symmetric or
antisymmetric Young tableaux will be idempotent with \(\alpha_\Y = 1\).

Diagrammatically \(P_{\Y_a}^2\) is simply \(P_{\Y_a}\) connected to
\(P_{\Y_a}\), hence it may be viewed as a set of \emph{outer} symmetry
operators connected by a set of \emph{inner} symmetry operators. Expanding all
the inner symmetry operators and using the uniqueness of the non-zero
connection between the symmetrizers and antisymmetrizers of the \YoungPOps, we
find that each term in the expansion is either 0 or a copy of \(P_{\Y_a}\). For
a Young diagram with \(s\) rows and \(t\) columns there will be a factor of
\(1/|\row_i|!\) (\(1/|\col_j|!\)) from the expansion of each inner
(anti)symmetrizer, so we find
\begin{eqnarray*}
  P_{\Y_a}^2 & = & \alpha_{\Y_a}^2 \btrackYb{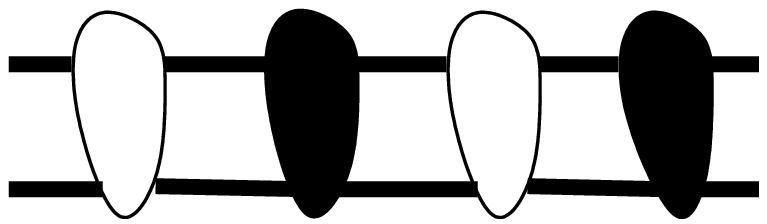}{12} \\ 
  & = & \frac{\alpha_{\Y_a}^2}
    {\prod_{i=1}^{s}|\row_i|! \prod_{j=1}^{t}|\col_j|!} 
    \sum_{\sigma} \btrackYb{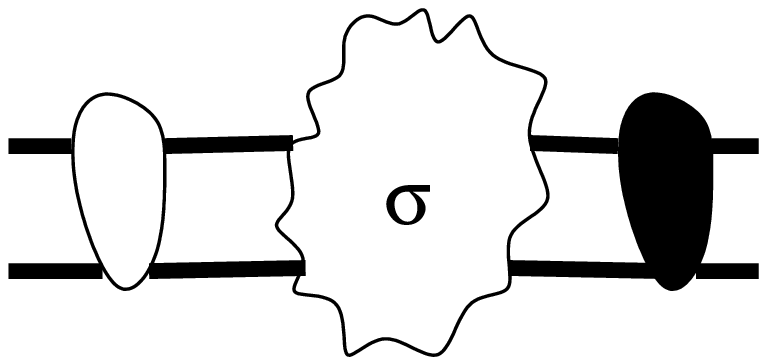}{12} \\
    & = & \alpha_{\Y_a}
      \frac{\kappa_\Y} {\prod_{i=1}^{s}|\row_i|!
        \prod_{j=1}^{t}|\col_j|!} \, P_{\Y_a} \,,
\end{eqnarray*}
where the sum is over permutation \(\sigma\) from the expansion of the inner
symmetry operators. Note that by the uniqueness of the connection between the
symmetrizers and antisymmetrizers, the constant \(\kappa_\Y\) is independent of
which tableau gives rise to the projection, and consequently the normalization
constant \(\alpha_\Y\) depends only on the Young diagram and not the tableau.

For a given \(\numBoxes\), consider the \YoungPOps\ \(P_{\Y_a}\) corresponding
to all the \(\numBoxes\)-box Young tableaux. Since the operators \(P_{\Y_a}\)
are orthogonal and in 1-1 correspondence with the Young tableaux, it follows
from the discussion in \refsect{grpalg} that there are no other operators of
\(\numBoxes\) lines orthogonal to this set. Hence the \(P_{\Y_a}\)'s form a
complete set, so that
\begin{equation}
  \id = \sum_{\Y_a} P_{\Y_a} \,.
  \label{compl}
\end{equation}
Expanding the projections the identity appears only once, so we have
\[P_{\Y_a} = \alpha_\Y \frac1{\prod_{i=1}^{s}|\row_i|! \prod_{j=1}^{t}
|\col_j|!} \left( \btrackYb{hSn.Sa}{4} + \dots \right) \,,\] and using this,
equation \refeq{compl} states
\begin{eqnarray}
  \btrackYb{hSn.Sa}{4} &=& 
    \left( k! \, \sum_{\Y} \frac{\alpha_\Y/|\Y|} 
      {\prod_{i=1}^{s}|\row_i|! \prod_{j=1}^{t}|\col_j|!} \right) 
      \,\btrackYb{hSn.Sa}{4} 
  \label{norm}
\end{eqnarray}
since all permutation different from the identity must cancel. When changing
the sum from a sum over the tableaux to a sum over the Young diagrams we use
that \(\alpha_\Y\) depends only on the diagram and that there are \(\Delta_\Y =
k!/|\Y|\) \(\numBoxes\)-standard tableaux for a given diagram. Choosing
\begin{equation}
  \alpha_\Y = \frac{\prod_{i=1}^{s}|\row_i|!
  \prod_{j=1}^{t}|\col_j|!}{|\Y|} \, ,
  \label{alpha}
\end{equation}
the factor on the right hand side of \refeq{norm} is 1 by \refeq{addition}.

Since the choice of normalization \refeq{alpha} gives the completeness relation
\refeq{compl}, it follows that it is also gives idempotent operators:
multiplying by \(P_{\Z_b}\) on both sides of \refeq{compl} and using
orthogonality, we find \(P_{\Z_b} = P_{\Z_b}^2\) for any Young tableau
\(\Z_b\).

\subsection{Dimensionality} \label{a_dim}

To prove the dimension formula \refeq{dimenUn} we need the identities
\begin{equation}
  \btrackYb{hSn.S}{5} = \frac1p
    \left( \btrackYb{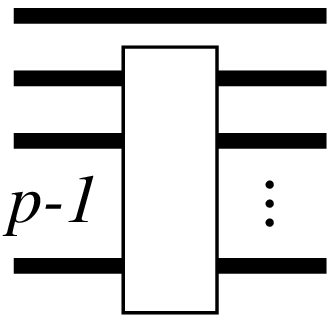}{5} + (p-1) \btrackYb{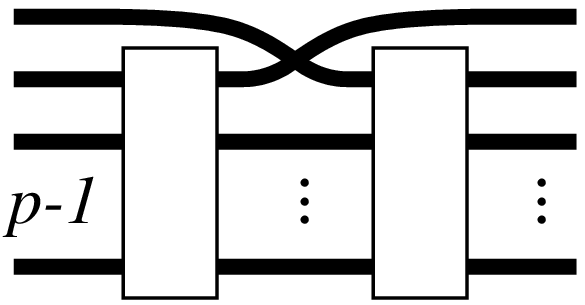}{9} \right) 
  \label{smart-S}
\end{equation}
and
\begin{equation}
  \btrackYb{hSn.A}{5} = \frac1p 
    \left( \btrackYb{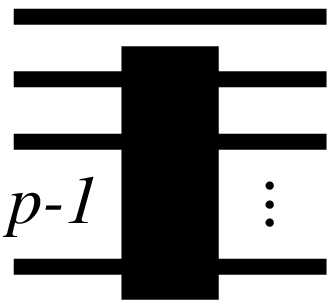}{5} - (p-1) \btrackYb{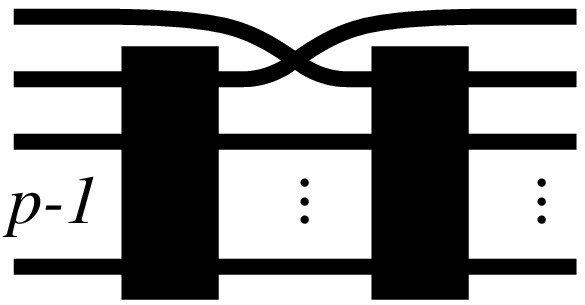}{9} \right)
  \label{smart-A}
\end{equation}
given in \refref{PCgr}. For Young tableaux with a single row or column, the
dimension formula can be derived directly using the relations \refeq{smart-S}
and \refeq{smart-A}.

Let \(\Y\) be a standard tableau with \(\numBoxes\) boxes, and \(\Y'\) the
standard Young tableau obtained from it by removing the box
containing~\(\numBoxes\). Draw the {\YoungPOps} corresponding to \(\Y\) and
\(\Y'\) and note that \(P_\Y\) with the ``last'' line traced is proportional
to~\(P_{\Y'}\).

Quite generally the contraction of the last line will look like
\begin{equation}
  \btrackYb{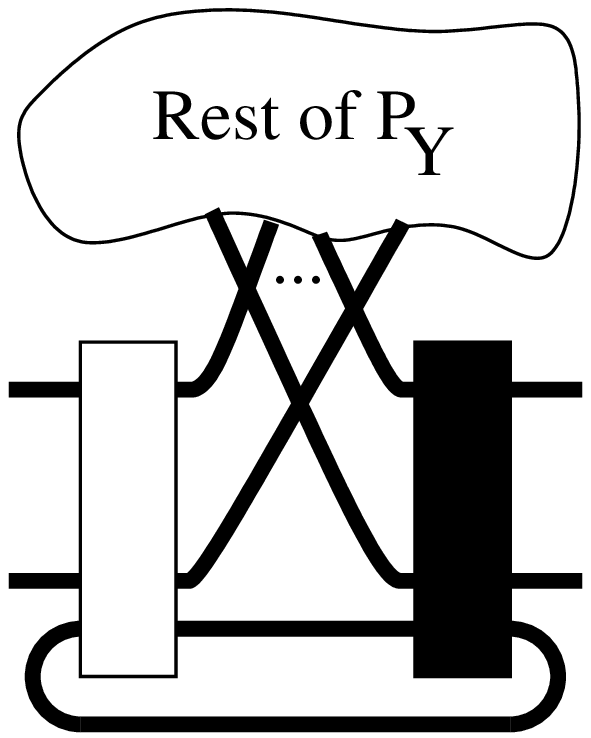}{9}
  \label{dcon}
\end{equation}
Using \refeq{smart-S} and \refeq{smart-A} we have \[\btrackYb{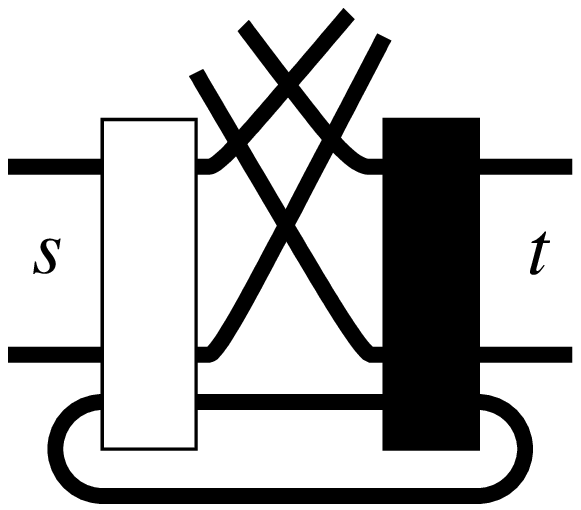}{9} =
\frac1s \left(\btrackYb{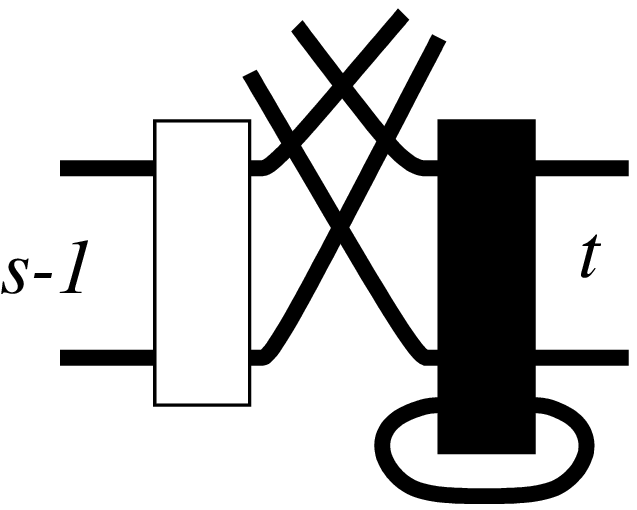}{10} + (s-1)\; \btrackYb{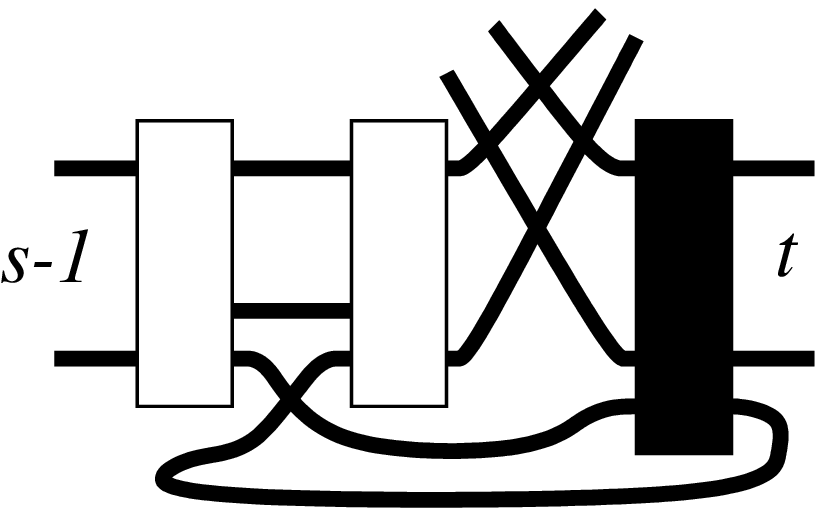}{13} \right)\]
\begin{eqnarray*}
    \quad &=& \frac{(n-t+1)}{st}\; \btrackYb{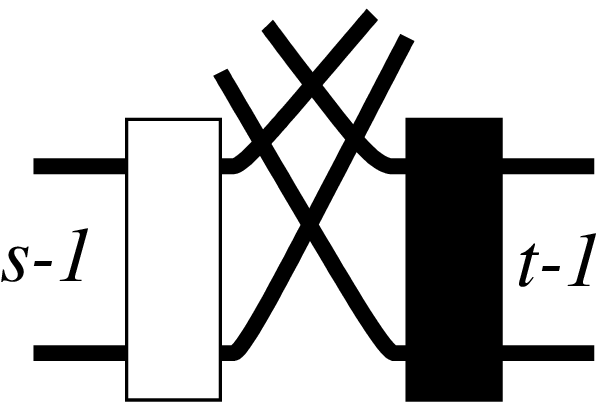}{10} + 
      \frac{(s-1)}{st}\;  \btrackYb{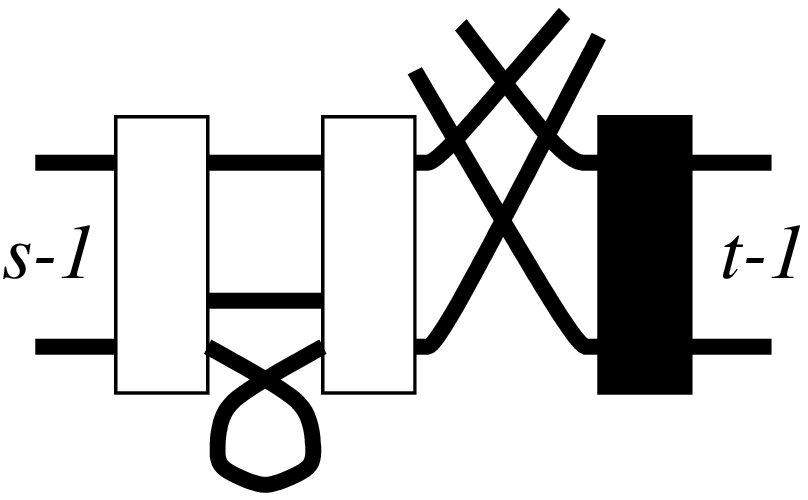}{13} \\
    \quad && \quad\qquad - \frac{(s-1)(t-1)}{st}\;  \btrackYb{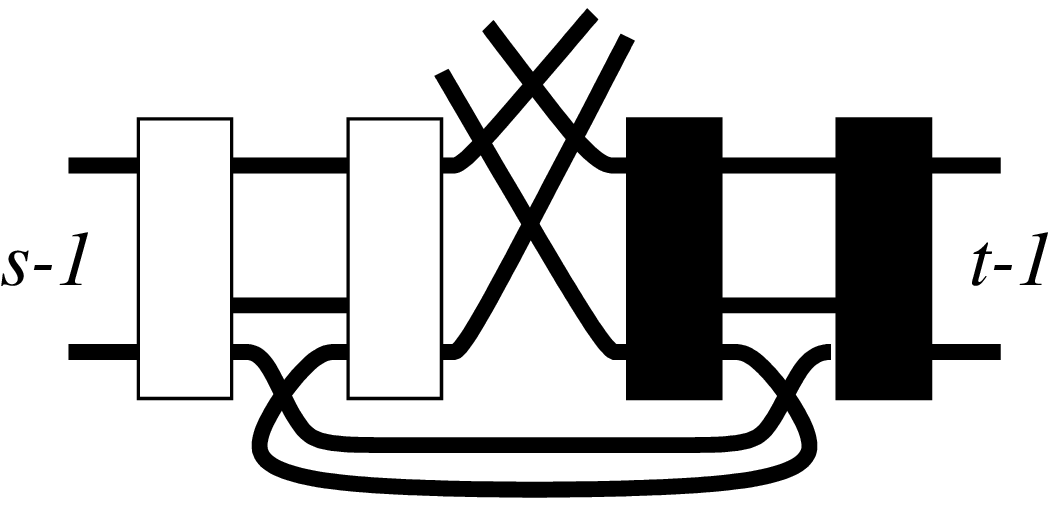}{17} \\
    \quad &=&  \frac{n-t+s}{st}\; \btrackYb{hYdim2d}{10} \\
    \quad && \quad\qquad\quad - \frac{(s-1)(t-1)}{st}\;
      \btrackYb{hYdim2f}{17} \,.
\end{eqnarray*}
Inserting this into \refeq{dcon} we see that the first term is proportional to
the projection operator \(P_{\Y'}\).

The second term vanishes: \[\btrackYb{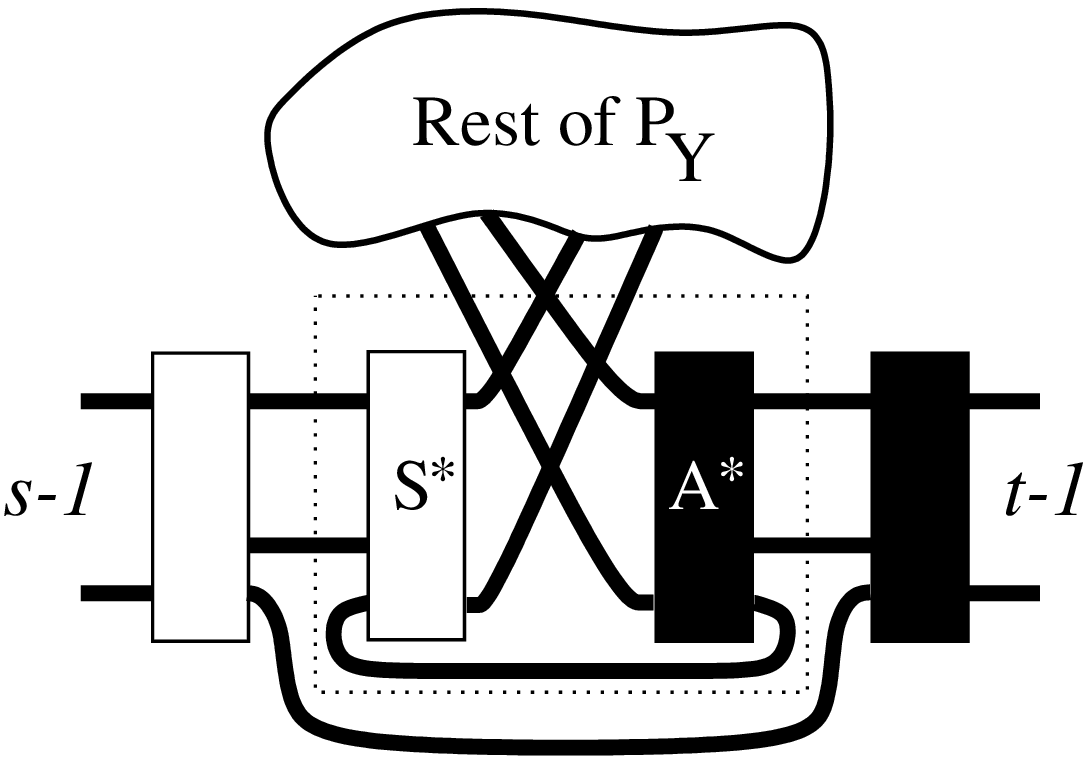}{17}\;= 0 \,.\] If we ignore the
internal structure within the dotted box we see that this is exactly of the
form of \(P_{\Y'}\), except that the ``last'' symmetrizer and antisymmetrizer
are connected by a line. There is a unique non-vanishing way of connecting the
symmetrizers and antisymmetrizers in \(P_{\Y'}\), and the ``last'' symmetrizer
and antisymmetrizer are not connected in this, as they correspond to a row and
column with no common box in the Young tableau. Therefore every term obtained
from the expansion of the dotted box must vanish.

The dimensionality formula follows by induction on the number of boxes in the
Young diagrams with the dimension of a single box Young diagram being \(n\).
Let \(\Y\) be a Young diagram with \(p\) boxes. We assume that the
dimensionality formula is valid for any Young diagram with \(p-1\) boxes. With
\(P_{\Y'}\) obtained from \(P_\Y\) as above, we have (using the above
calculation and writing \(D_\Y\) for the diagrammatic part of \(P_\Y\)):
\begin{eqnarray*}
  \dim P_\Y &=& \alpha_\Y \tr D_\Y = \frac{n-t+s}{st} \alpha_\Y \tr D_{\Y'} \\
  &=& (n-t+s) \alpha_{\Y'}\frac{|\Y'|}{|\Y|} \tr D_{\Y'} \\
  &=& (n-t+s)\frac{f_{\Y'}}{|\Y|} = \frac{f_{\Y}}{|\Y|}
\end{eqnarray*}
This completes the proof of the dimensionality formula~\refeq{dimenUn}.

\bibliography{groups-bibliography}
\bibliographystyle{adk-h-physrev}

\end{document}